\documentclass[sigconf]{acmart}
\AtBeginDocument{%
  }

\renewcommand\footnotetextcopyrightpermission[1]{} 

\usepackage{xspace}
\usepackage{graphicx}
\usepackage{amsmath,amsfonts}
\usepackage{graphicx}
\usepackage{textcomp}
\usepackage{xcolor}
\usepackage{listings}
\usepackage{enumitem}
\usepackage{microtype}
\usepackage{caption}
\usepackage{subcaption}
\usepackage{booktabs}
\usepackage[ruled,vlined,linesnumbered]{algorithm2e}
\usepackage{setspace}
\usepackage[most]{tcolorbox}
\usepackage{bbding}
\usepackage{multirow}
\usepackage{multicol}
\usepackage{colortbl}
\usepackage{array}
\usepackage{tabularx}
\usepackage{booktabs}
\usepackage{arydshln}
\usepackage{titlesec}

\tcbset{
    myfindingbox/.style={ 
        colback=gray!10, 
        colframe=white, 
        boxrule=0pt, 
        arc=1mm, 
        outer arc=1mm, 
        left=1mm, right=1mm, top=1mm, bottom=1mm, 
    }
}

\begin{document}
\newcommand{\dataset}{ContextCRBench\xspace}
\newcommand{\moduleA}{raw data crawling module\xspace}
\newcommand{\moduleB}{comprehensive context extraction module\xspace}
\newcommand{\moduleC}{multi-stage data filtering module\xspace}

\newcommand{\wxc}[1]{\textcolor{brown}{{#1}}}
\newcommand{\wangxc}[1]{\textcolor{blue}{{#1}}}
\newcommand{\yun}[1]{\textcolor{magenta}{#1}}

\titlespacing{\section}{0pt}{1pt}{1pt}
\titlespacing{\subsection}{0pt}{1pt}{1pt}
\titlespacing{\subsubsection}{0pt}{1pt}{2pt}

\title{Benchmarking LLMs for Fine-Grained Code Review with Enriched Context in Practice}









\author{Ruida Hu$^1$, Xinchen Wang$^1$, Xin-Cheng Wen$^1$, Zhao Zhang$^2$}
\author{Bo Jiang$^2$, Pengfei Gao$^2$, Chao Peng$^2$, Cuiyun Gao$^1$}

\affiliation{
  \institution{$^1$ Harbin Institute of Technology, Shenzhen, China; $^2$ ByteDance, Beijing, China}
  \country{}
}
\email{
200111107@stu.hit.edu.cn, 
200111115@stu.hit.edu.cn, 
xiamenwxc@foxmail.com, 
zhangzhao.a@bytedance.com, 
}
\email{
jiangbo.jacob@bytedance.com,
gaopengfei.se@bytedance.com, 
pengchao.x@bytedance.com,
gaocuiyun@hit.edu.cn
}

\settopmatter{printacmref=false, printccs=false, printfolios=false}
\renewcommand\footnotetextcopyrightpermission[1]{} 



\begin{abstract}
Code review is a critical practice for ensuring software quality in modern software development, and the recent advancements in Large Language Models (LLMs)
have demonstrated efficacy in facilitating automated code review processes.
However, existing benchmarks for code review have the following limitations.
\textbf{(1)} They lack the rich semantic context. Current benchmarks often provide 
code changes and fail to incorporate key textual information such as issue descriptions,
which are essential for understanding the intent behind a code change.
\textbf{(2)} 
They frequently exhibit data quality issues due to the absence of rigorous validation mechanisms during the curation process.
This negligence results in the incorporation of noisy entries, 
such as reviews on outdated code,
ultimately resulting in unreliable model evaluation.
\textbf{(3)} Most existing benchmarks operate at a file or commit level, failing to evaluate the fine-grained, line-level analysis essential for precise code understanding.
To address the limitations of existing datasets, we present \dataset, a high-quality, context-rich benchmark designed for fine-grained evaluation of LLMs in code review tasks.

Our construction pipeline consists of three main modules.
First, the \textbf{Raw Data Crawling module} collects over 153.7k issues and PRs from selected top-tier repositories.
Next, the \textbf{Comprehensive Context Extraction module} establishes rich context by rigorously linking issue-PR pairs for textual context and extracting the full surrounding function or class for code context.
Finally, our \textbf{Multi-stage Data Filtering module} applies a series of checks to remove entries that are outdated, improperly formatted, or identified as low-value by an LLM-based classifier.
This rigorous process yields the final benchmark of 67,910 entries.
Each entry in our benchmark is enriched with both textual context
and code context.

We design our benchmark to support three core evaluation scenarios aligned with the code review lifecycle:
(1) hunk-level quality assessment, assessing if a given diff hunk needs further review; (2) line-level defect localization, identifying the specific lines within the diff hunk needed to comment; and (3) line-level review comment generation, generating an actionable comment for an identified code line.
Leveraging \dataset, we conduct a comprehensive evaluation of eight popular LLMs, including four leading closed-source and four open-source models. 
We find that current LLMs still exhibit great limitations in code review, and the textual context often yields greater performance improvements than providing only the surrounding code context.
In an industrial application at ByteDance, \dataset serves as the core reward signal for a self-evolving code review tool, guiding it to a 61.98\% relative performance improvement.
This validates the practical utility and effectiveness of our benchmark in industrial applications.
All resources are available at \url{https://github.com/kinesiatricssxilm14/ContextCRBench}.
\end{abstract}

\maketitle

\section{Introduction}

\begin{figure}[t]
	\centering
	\includegraphics[width=.47\textwidth]{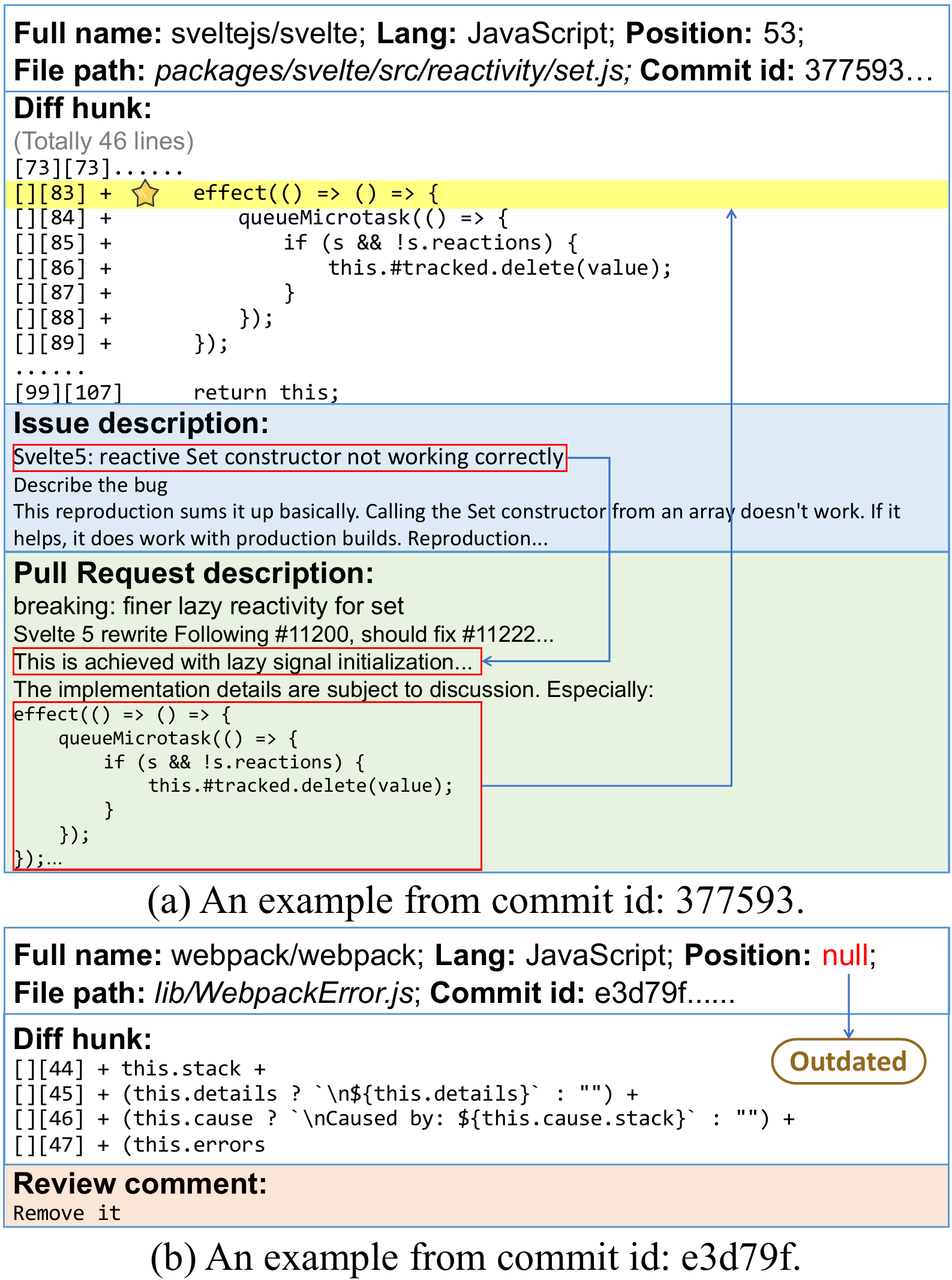}
    \vspace{-1em}
    \caption{Examples of the diff hunk for illustrating the challenges of current code review benchmarks. (a) illustrates a high-quality entry where red boxes highlight the issue (problem), the PR (solution), and the corresponding diff hunk (the code change). The starred line of code indicates the actual location of the review comment.
    }
\label{fig:intro_example}
\vspace{-1.8em}
\end{figure}

In recent years, Large Language Models (LLMs) have been proven effective in numerous domains.
~\cite{llm1, llm2, llm3}, 
leading to the emergence of numerous powerful Large Code Models (LCMs). These models have been increasingly adopted to automate different aspects of the software development lifecycle, including code completion, code generation, and code review~\cite{10.1145/3510003.3510621,10.1145/3540250.3549081, yin2023automatic, sawant2023code}. 
Code review is a critical practice in modern software development, playing a pivotal role in ensuring
code quality, identifying defects, and facilitating knowledge sharing among developers~\cite{survey1, survey2, survey3, survey4, survey5}.  
To understand and benchmark the capabilities of LLMs in code review, several benchmarks and evaluation frameworks have been proposed~\cite{10.1145/3540250.3549081,9825760,7883306,9896360}. 
These early efforts provide crucial references and have significantly advanced the field of automated code review.

\definecolor{darkgreen}{rgb}{0,0.5,0}
\definecolor{lightblue}{RGB}{230,240,255}
\newcommand{\cross}{\textcolor{red}{\textbf{\XSolidBrush}}}
\newcommand{\tick}{\textcolor{darkgreen}{\Checkmark}}

\newcolumntype{C}{>{\centering\arraybackslash}X}
\newcolumntype{M}[1]{>{\centering\arraybackslash}m{#1}}
\definecolor{deepgreen}{RGB}{0,100,0}
\setlength{\tabcolsep}{3.8pt} 
\renewcommand{\arraystretch}{0.1}

\begin{table}[t]
\scriptsize
\centering
\caption{Comparison of code review benchmarks. Our \dataset is the first to provide rich semantic context and support a fine-grained evaluation. QE, RCG, CR, DL, and RR indicate tasks of quality estimation, review comment generation, code refinement, defect localization, and reviewer recommendation, respectively. ``-'' means unkown.}
\vspace{-1.5em}
\label{tab:compare_dataset}
\footnotesize
\begin{tabular}{M{1.8cm}M{0.7cm}M{1.16cm}M{1.20cm}M{1.6cm}M{0.5cm}}
\toprule
\rowcolor{gray!20} \textbf{Benchmark} & \textbf{\#Langs} & \textbf{Context} & \textbf{Granularity} & \textbf{Filtering} & \textbf{Task}\\
\midrule
Li et al.~\cite{10.1145/3540250.3549081} & 9 & - & hunk & repo select, comment clean & QE RCG CR \\
\midrule
CORMS~\cite{10.1145/3540250.3549115} & - & metadata & hunk & bot remove, reviewer filter & RCG RR\\
\midrule
CORRECT~\cite{7883306} & 3 & PR Desc. & PR & - & RR\\
\midrule
CROP~\cite{paixao2018crop} & - & metadata & hunk & - & QE RCG CR\\
\midrule
Rong et al.~\cite{9793876} & - & metadata, PR Desc. & PR & bot remove, reviewer filter & RCG RR\\
\midrule
Tufano et al.~\cite{10.1145/3510003.3510621} & 1 & comment text & function & length filter & RCG CR\\
\midrule
Hong et al.~\cite{9825760} & 2 & - & line, \quad \quad file & reviewer filter, non-source file remove & DL \\
\midrule
\rowcolor{lightblue}
\textbf{\dataset} & 9 & metadata, issue and PR Desc., code before, code after & \textbf{line}, \quad hunk, \quad function & repo select,\quad bot remove, issue-PR pair, outdated diff filter, LLM-based filter ... & QE DL RCG \\
\bottomrule
\end{tabular}
\vspace{-2em}
\end{table}
\renewcommand{\arraystretch}{1.0}

However, despite this progress, a gap remains between current benchmarking practices and the complexities of real-world code review. Existing benchmarks still suffer from three fundamental limitations that we identify and address:

\textbf{(1) Lack of rich semantic context:} 
In real-world scenarios, code changes are often driven by specific requirements, such as bug fixes or new feature implementation, which are typically documented in issues and pull requests (PRs). However, most existing benchmarks overlook this crucial contextual information.
For example, as shown in Figure~\ref{fig:intro_example} (a), reviewing only the diff hunk in the left part often leads to comments focused on code style or syntax, while missing the true intent of the code change. In contrast, the issue and PR descriptions in the right part highlight the underlying bug and proposed fix, respectively. This context enables reviewers to focus on new lines 83 to 89 of the diff hunk and provide meaningful, functionality-oriented comments.
As a result, LLMs are forced to infer the intent behind code changes without truly understanding the justification. 
\textbf{(2) Poor data quality:}
Many existing benchmarks suffer from poor data quality that can lead to unreliable model evaluation.
Figure~\ref{fig:intro_example} (b) shows the metadata of an outdated diff hunk, including attributes such as ``\texttt{Position}'' and ``\texttt{Diff hunk}''. Specifically, the ``\texttt{Position}'' attribute is null, indicating that the comment targets an outdated or deleted diff hunk. Besides, the ``\texttt{Reviewer comment}'' simply states ``Remove it'', lacking any rationale or alternative solution. Such outdated and low-quality data fail to reflect typical code review practices.
Including such outdated and low-quality data fails to reflect realistic code review scenarios and hinders the accurate assessment of LLM capabilities.
\textbf{(3) Lack of fine-grained and line-level analysis:} 
Existing benchmarks typically operate at the file and hunk level, failing to support the fine-grained, line-level evaluation necessary for generating specific and localized review feedback.
As shown in Figure~\ref{fig:intro_example} (a), the diff hunk spans 46 lines of code. Generating review comments for the entire hunk often results in unfocused feedback that offers limited value to developers. In practice, the reviewer comments specifically on line 83, providing targeted and fine-grained feedback and helping accelerate the code review cycle.
Although Hong et al.~\cite{9825760} also focus on line-level analysis, their evaluation is framed as a binary classification task—determining whether a given line requires a comment. This setup does not directly correspond to our defect Localization task, which requires a model to proactively identify the specific lines containing defects within a larger code change.
Consequently, there is still no standardized way to assess LLMs' ability to perform a comprehensive, end-to-end review at the line level—from precisely localizing problematic lines (``Localization'') to generating actionable, human-like feedback (``Comment Generation'').
\begin{figure}[t]
	\centering
	\includegraphics[width=.5\textwidth]{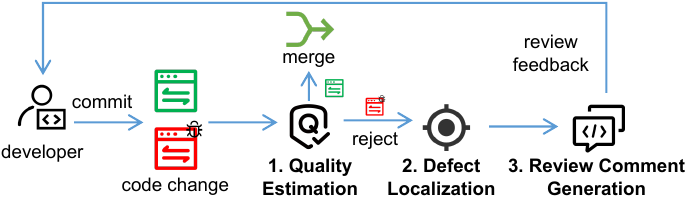}
    \vspace{-1.5em}
    \caption{The automated code review workflow and our three evaluation tasks.}
    \vspace{-2em}
\label{fig:review_process}
\end{figure}

To address these limitations, we introduce \dataset, a novel, large-scale benchmark designed for comprehensive, fine-grained, and context-aware evaluation of LLMs in code review. We construct \dataset through a meticulous filtering pipeline applied to over 2.7 million issues and PRs from 90 popular repositories across nine programming languages, resulting in a final dataset of 67,910 high-quality samples. Each sample in \dataset precisely links a line-level review comment to its code change and, crucially, is enriched with both textual context (issue and PR descriptions) and code context (the full before-and-after function or class).
We summarize the key differences between \dataset and prior code review datasets in Table~\ref{tab:compare_dataset}.

To facilitate a comprehensive assessment, we propose to evaluate models on three distinct tasks that simulate a realistic review process, as illustrated in Figure~\ref{fig:review_process}. These tasks are: (1) Hunk-level Quality Estimation, to assess the overall quality of a change; (2) Line-level Defect Localization, to identify the specific lines of concern; and (3) Line-level Review Comment Generation, to provide review feedback.
Based on \dataset, we conduct an extensive evaluation of eight popular LLMs (four closed-source and four open-source) on these three tasks, systematically investigating the impact of different context combinations on their performance. Our experiments lead to several key findings. For example, we find that current LLMs still fall short of the requirements for reliable, automated code review. We also find that enriching prompts with context generally enhances performance, with textual context such as issue and PR descriptions providing a greater boost than the code context of surrounding functions.
In addition, we apply \dataset in an industrial application at ByteDance to guide a self-evolving code review tool. It is used as the reward signal and achieves a 61.98\% relative performance improvement.

In summary, this paper makes the following contributions:
\begin{enumerate}

\item We introduce \dataset, the first large-scale, multi-language benchmark that provides rich semantic context (both textual and code) for fine-grained code review tasks.
\item We conduct an in-depth analysis of eight LLMs, revealing their current capabilities and limitations in context-aware, fine-grained code review and providing valuable insights.
\item We demonstrate the practical value of \dataset in a real-world industrial application at ByteDance, where it serves as the reward signal to guide a self-evolving code review tool to a 61.98\% relative performance improvement.
\end{enumerate}


\section{Approach}
\label{approach}
\begin{figure}[t]
	\centering
	\includegraphics[width=.48\textwidth]{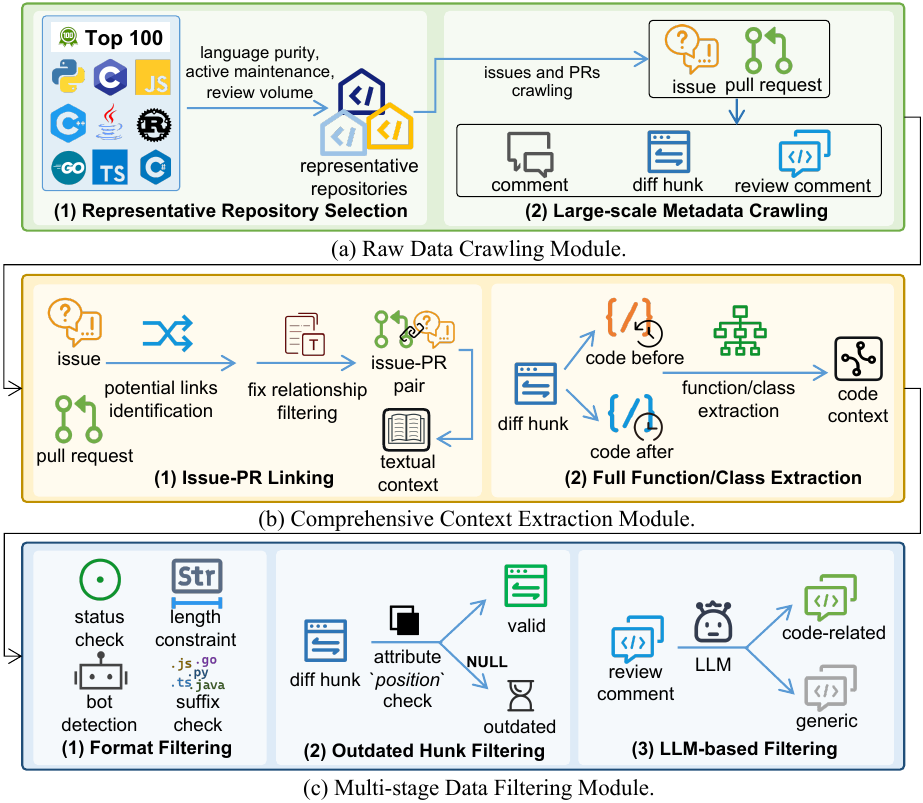}
    \vspace{-2em}
    \caption{The pipeline of \dataset construction. It consists of three main modules: \moduleA for collecting large-scale issues and PRs, \moduleB for constructing rich textual and code context, and \moduleC for filtering noisy data and producing the final high-quality benchmark.}
    \vspace{-1.5em}
\label{fig:build_framework}
\end{figure}

The overall procedure of our data construction pipeline is illustrated in Figure~\ref{fig:build_framework}. The pipeline consists of three main modules: (1) \moduleA, responsible for collecting a large amount of issues\footnote{Issues refer not only to bugs but also to requests for new features in GitHub.} and PRs to construct the initial dataset; (2) \moduleB, designed to build rich textual and code context for each diff hunk; and (3) \moduleC, which applies multi-layered filtering to remove noise and produce the final high-quality, fine-grained benchmark.

\subsection{Raw Data Crawling Module}
To ensure high-quality, large-scale, and broad coverage of real-world scenarios, the \moduleA first involves selecting a diverse set of actively maintained repositories across nine major programming languages, and then systematically collecting all issues, PRs, and associated comments from these repositories to construct the initial dataset.

\subsubsection{Representative Repository Selection.}
To build a high-quality and diverse benchmark, we start by
selecting nine widely-adopted programming languages~\cite{topprogramminglanguage}, including Python, Java, JavaScript, C, C++, C\#, Rust, TypeScript, and Go. For each language, we then choose repositories from a list of top-starred GitHub repositories based on the following rigorous criteria\footnote{The crawling was performed in April 2025.}:

$\bullet$ Popularity: The repository should be among the top 100 most-starred~\cite{top100} for its language, because they often have well-organized issues and PRs.

$\bullet$ Language purity: The target language should constitute over 60\% of the repository for language-centric review scenarios.

$\bullet$ Active maintenance: The repository must show development activity within the last six months to reflect current and relevant development practices.

$\bullet$ Review volume: The repository should have a cumulative total of over 5,000 issues and PRs to provide a large pool to sample from.

Following this series of filtering steps, we further conduct a manual check on the remaining projects to ensure that only well-maintained and representative projects are included. Ultimately, we curate 90 high-quality projects for \dataset.
\subsubsection{Large-scale Metadata Crawling.}
To ensure complete contextual details, we perform a large-scale metadata crawl to gather the raw data. Specifically, we utilize the GitHub REST API, a powerful interface for accessing public repository data, to systematically retrieve all issues, PRs, and their associated comments from the selected repositories.
To focus on recent data, we exclusively collect issues and PRs created since January 2021 to mitigate the risk of overlap with the models' pre-training corpora. This initial collection yields a massive corpus of 153.7k issues and PRs, which serves as the basis for subsequent context construction and data filtering.

\subsection{Comprehensive Context Extraction Module}

To address the limitations of existing benchmarks that lack rich semantic context, we enhance each diff hunk with comprehensive project-level context during the \moduleA.
This module aims to replicate the information a human reviewer would use by sourcing both textual and code context.

\subsubsection{Textual Context: Linking Issues to PRs.}
\label{subsub:crossref}

The textual context explains the 
reason for the code change issue, which
often appears in the associated issue and PR. 
To accurately identify pairs where a PR is to fix a specific issue, we design a two-step filtering process.

\textbf{Step 1: Identifying Potential Issue-PR Links.}
First, we identify all potential links between issues and PRs. Based on the GitHub documentation~\cite{git3}, a developer can reference an issue in a PR description using various formats, such as a full URL, a shorthand notation (e.g., ``\texttt{\#26}''), or a cross-repository reference (e.g., ``\texttt{owner/repo\#26}''). We begin by using pattern matching to parse the PR's title, body, and comments to find all such references, creating a candidate pool of potentially related issue-PR pairs.
However, a simple textual reference does not guarantee a true, functional link within GitHub's system. To confirm the relationship with high fidelity, we then query the GitHub GraphQL API~\cite{git2} for each candidate pair. We verify that a timeline event explicitly connecting the issue and the PR exists, thus confirming an authentic association and filtering out coincidental or incorrect textual matches.

\textbf{Step 2: Filtering for Fix Relationships.}
Not all confirmed links represent a ``fix'' relationship; some might be simple mentions or discussions. To isolate the pairs where the PR is specifically intended to resolve the issue, we perform a final keyword check. Following GitHub's standard practice, we search the PR description for specific keywords that trigger an automatic link-and-close action, such as ``\texttt{closes}'', ``\texttt{fixes}'', or ``\texttt{resolves}'' (e.g., ``\texttt{closes \#10}''). Only the issue-PR pairs that contain these explicit ``fix'' keywords are retained as our high-quality textual context.

\begin{algorithm}[t]
    \footnotesize
    \setstretch{0.3}
    \SetAlgoLined
    \footnotesize
    \caption{Textual Context Extraction Algorithm}
    \label{alg:alg1}
    \KwIn{$issue$;}
    \KwOut{$fixingPRList$;}

    $referenceList \gets \emptyset$;

    
        $content \gets entry.title + entry.body + entry.comments$;

    \tcp{Extract the candidate PRs}
    $candidateList \gets \texttt{ExtractCandidates}(content)$; 

    \ForEach{$candidate \in candidateList$}{
        $validReferenceList \gets \texttt{ExtractInboundReferences}(candidate)$; 

        \If{$entry \in validReferenceList$}{
            $referenceList.\texttt{add}(candidate)$; 
        }
    }

    \tcp{Further filter PRs exactly fixing the given issue}
    $fixingPRList \gets \texttt{KeywordCheck}(referenceList)$;

    \Return{$fixingPRList$;}
\end{algorithm}

As detailed in Algorithm \ref{alg:alg1}, we take a GitHub issue as input and return all PRs fixing it. In Line 2, we gather the content for the issue. In Line 3, we extract all references using the matching patterns. In Lines 4-9, we use GitHub GraphQL to sequentially query ``\textit{CrossReferenceEvent}'' for each candidate. For each candidate, we check whether it is referenced by the input issue (Line 6). If the candidate is found in the $validReferenceList$, it indicates a confirmed reference, and the candidate is added to the output list (Line 7). Finally, we further identify the exact fixing relationships by checking the keywords (Line 10).

This two-step process allows us to comprehensively identify issue-PR pairs. For each validated pair, we extract the full description text (title and body) from both the issue and the PR, forming the complete textual context for the associated code changes.

\subsubsection{Code Context: Full Function/Class Extraction.}

A diff hunk is often a fragment of a larger logical unit, so to provide this crucial structural information, we extract the complete code context surrounding each commented diff hunk. Since a diff hunk often represents only a fragment of a larger logical unit, we extract the complete code context surrounding each commented diff hunk to provide essential structural information in the project.
We first identify the target file path and the specific commit SHAs for both the before and after states of the code changes, corresponding to the base and head commits of the change, respectively. We then utilize the GitHub REST API to retrieve the full content of the source file at each of these two versions. With the complete file content, we employ Tree-Sitter~\cite{Tree-sitter} to build the AST.
By mapping the line numbers of the diff hunk to the AST, we retrieve the source code of the smallest enclosing scope, typically a function, method, or class, according to the code syntax of different programming languages. If a diff hunk constitutes an entire file addition or deletion, the corresponding ``before'' or ``after'' context may be empty.

\subsection{Multi-stage Data Filtering Module}
Raw data collected from software repositories is inherently noisy. To address the limitations of existing benchmarks that lack meticulous filtering steps and ensure the high quality and reliability of \dataset, we apply a multi-stage filtering process, including format, outdated hunk, and LLM-based filtering.
\subsubsection{Format Filtering.}
We first apply a set of format rules to remove clearly invalid or irrelevant entries.
This step ensures that the collected information is comprehensive, appropriately scoped in length, and devoid of duplicate entries. The details are as follows:

$\bullet$ Status Check: We retain only PRs that have been merged and issues that are closed. This ensures that our data represents completed and resolved work items.

$\bullet$ Length Constraints: We remove samples where the code changes or the review comments are excessively long (exceeding 1MB). This helps to eliminate outliers and focuses on typical review scenarios.


$\bullet$ Bot Removal: We remove entries where review comments are generated by bots, ensuring that our benchmark reflects genuine human code review interactions.
 
$\bullet$ Suffix Check: We check the suffix of the file, and remove entries whose code changes involve only documentation or non-functional files, such as ``\texttt{.md}'' or ``\texttt{.rst}'' files.

\subsubsection{Outdated Hunk Filtering.}
In a typical PR lifecycle, code undergoes multiple revisions. Consequently, review comments made on an early commit may become outdated if the commented code lines are substantially modified or deleted in subsequent commits within the same PR. 
Such comments are no longer relevant to the final state of the code change and introduce much noise.
We identify outdated diff hunks by checking their ``\texttt{position}'' attribute retrieved via the GitHub API; a ``\texttt{Null}'' or invalid value indicates that the comment can no longer be mapped to the final diff. By removing these entries, we ensure that every diff hunk in \dataset represents a valid interaction between a reviewer and the code that is ultimately considered for merging.

\begin{table}[t]
\centering
\caption{Statistics of \dataset for each task. QE, DL, RCG, Lang\#, DH LOC\#, TC Length, CC LOC\# indicate quality estimation, defect localization, review comment generation, number of programming languages, total lines of code in diff hunks, total length of textual context, total lines of code context, respectively.}
\vspace{-1em}
\label{statistic}
\footnotesize
\resizebox{\columnwidth}{!}{
\begin{tabular}{cccccc}
\toprule
\rowcolor{gray!20} \textbf{Task} & \textbf{Entry\#} & \textbf{Lang\#} & \textbf{DH LOC\#} & \textbf{TC Length} & \textbf{CC LOC\#} \\
\midrule
QE & 67,910 & 9 & $\sim$6.40M & $\sim$275.17M & $\sim$148.37M \\
DL & 13,512 & 9 & $\sim$1.00M & $\sim$51.12M & $\sim$28.14M \\
RCG & 13,512 & 9 & $\sim$1.00M & $\sim$51.12M & $\sim$28.14M \\
\bottomrule
\end{tabular}
}
\vspace{-2em}
\end{table}

\subsubsection{LLM-based Filtering.}

LLM-as-a-judge for filtering is widely used in natural language generation and code intelligence tasks~\cite{bavaresco2025llmsinsteadhumanjudges, codejudge, ice-score}.
To ensure the benchmark contains only semantically meaningful entries, we employ advanced LLM (e.g., GPT-4o) to assess the quality of each review comment based on its relevance to the associated code. Our prompt design incorporates few-shot learning~\cite{sung2018learning} by providing the model with clear examples of high-quality (specific, actionable feedback) and low-quality (e.g., ``LGTM'', generic conversation) comments. We also leverage Chain-of-Thought (CoT)~\cite{wei2022chain} prompting, instructing the model to first reason about the code's logic and the comment's intent before making a final judgment. This structured strategy enhances the model's decision-making accuracy. Entries flagged by the LLM as having low relevance or poor quality are discarded. We further validate the reliability of this filtering step through a manual audit. Three human experts, each with relevant programming experience of over 5 years and familiarity with the project repositories, independently evaluate a random sample of 300 LLM-judged entries. The results show over 85\% agreement between the LLM's decisions and the human expert consensus. The inter-annotator agreement among the experts is also high, with a Kappa coefficient over 90\%. This robust validation confirms that our semantically-aware filtering step ensures a high standard of textual quality for the entries in our benchmark.


To provide a comprehensive overview of our benchmark, we present a statistical summary in Table~\ref{statistic}. This table details the scale of each task's dataset, including the number of entries, the count of programming languages, and the total volume of diff hunks, textual context, and surrounding code context.

\subsection{Research Questions}
\subsubsection{RQ1: How do different LLMs perform in the three tasks of code review?} 
In this research question, we divide the code review task into three scenarios, evaluating and comparing the performance of LLMs in each scenario, given the basic contexts.
\subsubsection{RQ2: How do different types of contexts affect LLMs' performance in code review?}
In this research question, we investigate the influence of involving different types of contexts on the model performance in code review.
\subsubsection{RQ3: How does LLM performance in code review vary across different programming languages?}
In this research question, we investigate whether the performance of LLMs in code review is influenced by the programming language.

\section{Experimental Setup}
\label{setup}
\subsection{Selected LLMs}
To comprehensively evaluate modern LLMs, we select a diverse set of eight models, as detailed in Table~\ref{tab:selected_LLMs}. Our selection includes four leading popular closed-source models (GPT-4, GPT-4o, GPT-3.5-turbo, and Claude 3.5 Sonnet) and four widely-used powerful open-source models specifically optimized for code (Codestral and three variants of Qwen2.5-Coder-Instruct).

\subsection{Metrics}
\label{metric}
We adopt the following metrics for the three code review tasks:

\subsubsection{Hunk-level quality estimation.}
As a binary classification task, we apply F1-score, accuracy, and precision.

$\bullet$ \textbf{F1-score}: It is the harmonic mean of precision and recall, balancing the trade-off between them.

$\bullet$ \textbf{Accuracy}: It evaluates the proportion of correct predictions.

$\bullet$ \textbf{Precision}: It measures the proportion of predicted positive cases that are actually positive.

\renewcommand{\arraystretch}{0.4}

\begin{table}[t]
\setlength{\tabcolsep}{3.5pt} 
\scriptsize 
\centering
\caption{Comparison of selected LLMs. ``-'' means unknown.}
\vspace{-1em}
\label{tab:selected_LLMs}
\footnotesize
\resizebox{\columnwidth}{!}{
\begin{tabular}{ccccc}
\toprule
\rowcolor{gray!20} \textbf{LLMs} & \textbf{Size} & \textbf{Time} & \textbf{Open-source} & \textbf{Abbreviation} \\
\midrule
GPT-4 & - & 2023-03 & \cross & GPT-4\\
GPT-4o & - & 2024-05 & \cross & GPT-4o \\
GPT-3.5-turbo & - & 2023-05 & \cross & GPT-3.5 \\
Claude3.5-Sonnet & - & 2024-06 & \cross & Cl-S \\
Codestral & 22B & 2025-01 & \tick & Codestral \\
Qwen2.5-Coder-Instruct & 32B & 2024-11 & \tick & CQ-32B \\
Qwen2.5-Coder-Instruct & 14B & 2024-11 & \tick & CQ-14B \\
Qwen2.5-Coder-Instruct & 7B & 2024-09 & \tick & CQ-7B\\
\bottomrule
\end{tabular}
}
\vspace{-2em}
\setlength{\tabcolsep}{6pt} 
\end{table}

\renewcommand{\arraystretch}{1.0}

\renewcommand{\arraystretch}{0.8}

\begin{table*}[hbtp]

\scriptsize 
\centering
\caption{Results in the tasks of quality estimation, defect localization, and review comment generation. The largest and second-largest values in each row are highlighted in dark grey and light grey, respectively. The largest value is bold. $\sigma^2$ represents the variance of the data in this row. Cl-S refers to Claude3.5-Sonnet, and CQ refers to the Qwen-Coder series.}
\vspace{-1em}
\label{tab:base_result}
\setlength{\tabcolsep}{3.5pt} 
\footnotesize
\resizebox{0.8\textwidth}{!}{
\begin{tabular}{ccccccccccc}
\toprule
\rowcolor{gray!20} \textbf{Metric} & \textbf{GPT-3.5} & \textbf{GPT-4} & \textbf{GPT-4o} & \textbf{Cl-S} & \textbf{Codestral} & \textbf{CQ-32B} & \textbf{CQ-14B} & \textbf{CQ-7B} & \color{blue!80}\textbf{Avg} & \color{blue!80}\textbf{$\sigma^2$} \\
\midrule
\multicolumn{11}{c}{\textit{Quality estimation}} \\
\midrule
F1-score &  27.71 & 27.04 & 30.03 & \cellcolor{gray!50}\textbf{59.54} & \cellcolor{gray!30}53.68 & 21.51 & 53.21 & 33.78 & 38.31 & 189.96 \\
Accuracy & 22.22 & 21.96 & 24.07 & 44.63 & 41.67 & \cellcolor{gray!30}45.28 & 40.50 & \cellcolor{gray!50}\textbf{45.45} & 35.72 & 103.88 \\
Precision & 25.88 & 25.41 & 27.85 & \cellcolor{gray!50}\textbf{46.91} & \cellcolor{gray!30}44.51 & 38.03 & 43.82 & 42.98 & 36.92 & 72.45 \\
\midrule
\multicolumn{11}{c}{\textit{Defect localization}} \\
\midrule
Pass@1 & 15.74 & 16.48 & 15.83 & 17.59 & 17.59 & \cellcolor{gray!30}24.07 & 13.70 & \cellcolor{gray!50}\textbf{24.81} & 18.23 & 14.22 \\
Pass@3 & 34.63 & 34.26 & 33.06 & \cellcolor{gray!30}39.44 & 34.44 & \cellcolor{gray!50}\textbf{40.83} & 32.96 & 36.39 & 35.75 & 7.51 \\
Pass@5 & 45.09 & 44.81 & 43.43 & \cellcolor{gray!30}50.00 & 44.81 & \cellcolor{gray!50}\textbf{51.48} & 42.59 & 44.44 & 45.83 & 8.77 \\
Pass@10 & 59.81 & 61.67 & 59.54 & \cellcolor{gray!50}\textbf{67.04} & 59.72 & \cellcolor{gray!30}66.85 & 60.28 & 55.37 & 61.29 & 13.49 \\
\midrule
\multicolumn{11}{c}{\textit{Review comment generation}} \\
\midrule
Rouge-1 & 17.93 & \cellcolor{gray!30}18.11 & \cellcolor{gray!50}\textbf{18.16} & 16.69 & 12.95 & 16.20 & 16.22 & 13.43 & 16.21 & 3.62 \\
Rouge-L & 12.03 & \cellcolor{gray!30}12.10 &  \cellcolor{gray!50}\textbf{12.12} & 11.56 & 9.91 & 11.62 & 11.37 & 9.88 & 11.32 & 0.75 \\
Edit Similarity & \cellcolor{gray!30}21.34 & \cellcolor{gray!50}\textbf{21.40} & 21.28 & 20.38 & 16.99 & 19.76 & 19.76 & 18.61 & 19.94 & 2.08 \\
\bottomrule
\end{tabular}
}
\vspace{-1.5em}
\setlength{\tabcolsep}{6pt} 
\end{table*}

\renewcommand{\arraystretch}{1.0}

\subsubsection{Line-level defect localization.}
We evaluate the model's ability to identify defective lines within a given code diff precisely. We adopt the pass@k metric, where the LLM provides a ranked list of suspicious lines, ordered by confidence. A trial is considered successful if any of the top-k suggested lines match the ground-truth defective line. We report  ``\texttt{pass@1}'', ``\texttt{pass@3}'', ``\texttt{pass@5}'', and ``\texttt{pass@10}'' to measure the percentage of cases where a correct line is found within the top 1, 3, 5, and 10 predictions, separately. Among these, \texttt{pass@1} is the most stringent metric, representing the model's accuracy in identifying the correct line as its top suggestion.





\subsubsection{Line-level review comment generation.}
For review comment generation, we assess the quality of the generated text by comparing it against the ground-truth comment. We utilize the following metrics to measure semantic and lexical similarity:

$\bullet$ \textbf{ROUGE}: It measures the overlap between the generated comment and the reference comment. We specifically use two variants:

  (1) \textbf{ROUGE-1} focuses on the overlap of individual words (unigrams), evaluating how well the key terms are captured.

  (2) \textbf{ROUGE-L} measures the longest common subsequence, which assesses the structural and sentence-level similarity.
  
$\bullet$ \textbf{Edit Similarity}: This metric calculates the similarity based on the Levenshtein distance. It is computed as ``\texttt{1 - (edit\_distance / max\_length)}'', reflecting the normalized cost of transforming the generated text into the reference text through insertions, deletions, and substitutions. A higher score indicates a closer match.





\subsection{Implementation details}
To ensure deterministic and reproducible outputs from the LLMs, we uniformly set the temperature to 0 across all experiments. The prompt for each task is also available in our repository. Considering the financial costs, we construct a representative and balanced subset from our main dataset for our evaluation. Specifically, we first stratify the dataset by time, creating four temporal groups: 2021, 2022, 2023, and 2024 onwards. Within each of these four groups, we perform stratified sampling across the nine programming languages, randomly selecting 30 entries per language. This results in a final evaluation set of 1,080 entries. For the quality estimation task, we maintain a balanced distribution by sampling an equal number of merged and unmerged entries within each subgroup. For the defect localization and review comment generation tasks, we exclusively sample from unmerged entries, as these actions inherently occur during the phase before a pull request is merged.

\section{Experimental Result}
\label{result}
\subsection{RQ1: Model performance in code review}
To establish a performance baseline, we first evaluate the models using only the diff hunk, without providing additional code or textual context. The evaluation results are presented in Table~\ref{tab:base_result}. 

\textbf{(1) Despite recent advancements, current LLMs exhibit large limitations across all fundamental code review tasks.} As shown in Table~\ref{tab:base_result}, even the top-performing models struggle to achieve high scores. In quality estimation, the best F1-score is 59.54, indicating that LLMs still frequently misclassify the quality of code changes. The challenge is also pronounced in defect localization, where the top Pass@1 score is only 24.81. This means that in nearly three-quarters of cases, the model fails to identify the correct defective line on its first attempt. Similarly, for review comment generation, the Rouge-L scores, which measure structural similarity, are low (e.g., a maximum of 12.12), suggesting that the generated comments often lack the logical flow and precision of human-written feedback. This overall modest performance highlights that automated code review remains a highly challenging problem for the current generation of LLMs.

\textbf{(2) Specialized open-source code models demonstrate a competitive advantage in code-specific tasks.} While closed-source models like Claude3.5-Sonnet and GPT-4o excel in tasks requiring high-level classification or natural language fluency, the open-source models trained specifically on code show remarkable strength in more granular, code-centric tasks. This is most evident in defect localization, where the Qwen-Coder series (CQ-32B and CQ-7B) secures top ranks across multiple ``pass@k'' metrics. For example, CQ-7B's Pass@1 score of 24.81 is nearly 50\% higher than that of GPT-4o (15.83). This indicates that for tasks requiring deep code comprehension, such as identifying a single incorrect line, domain-specific models can be more effective than larger models.

\textbf{(3) Model performance variance reveals that 
quality estimation
is a more volatile task than 
defect localization and comment generation.}
The $\sigma^2$ column in Table~\ref{tab:base_result} highlights a critical difference in performance consistency. The quality estimation task
exhibits extremely high variance (e.g., 189.96 for F1-score). 
It indicates that the model faces challenges in precisely delineating which types of behaviors necessitate code review.
In contrast, defect localization and review comment generation show substantially lower variance (e.g., 14.22 for Pass@1 and 3.62 for Rouge-1). 
This implies that, for both tasks, the models demonstrate a better understanding of the task objectives. Consequently, the performance gap between different models on these tasks is substantially reduced.

\begin{tcolorbox}[myfindingbox]
    \textbf{Finding 1:} Current LLMs exhibit large limitations in automated code review.
    Specialized small LCMs are competitive in code-centric tasks like defect localization. The high performance variance in quality estimation suggests it is a particularly volatile and challenging task for LLMs.
\end{tcolorbox}

\renewcommand{\arraystretch}{0.8}

\definecolor{deepgreen}{RGB}{0,100,0}
\begin{table*}[hbtp]
\setlength{\tabcolsep}{3.5pt} 
\scriptsize 
\centering
\caption{Results in the quality estimation task. The largest and second-largest values in each column are highlighted in dark grey and light grey, respectively. The largest value is bold. Avg$_1$ and Improve$_1$ represent the average score and relative improvement rate for the four closed-source LLMs, respectively. Avg$_2$ and Improve$_2$ represent the average score and relative improvement rate for the four open-source LLMs. I, P, B, and A refer to issue description, pull request description, code before, and code after, respectively. The same highlighting and abbreviations are also used in Tables~\ref{tab:task2_result} and~\ref{tab:task3_result}.}
\vspace{-1em}
\label{tab:task1_result}
\footnotesize
\resizebox{\textwidth}{!}{
\begin{tabular}{cccccccccccccc}
\toprule
\rowcolor{gray!20}\textbf{Context} & \textbf{Method} & \textbf{GPT-3.5} & \textbf{GPT-4} & \textbf{GPT-4o} & \textbf{Cl-S} & \color{blue!80}\textbf{Avg$_1$} & \color{blue!80}\textbf{Improve$_1$} & \textbf{Codestral} & \textbf{CQ-32B} & \textbf{CQ-14B} & \textbf{CQ-7B} &  \color{blue!80}\textbf{Avg$_2$} & \color{blue!80}\textbf{Improve$_2$}\\
\midrule
\multicolumn{14}{c}{\textit{F1-score}} \\

\midrule
-& Base & 27.71 & 27.04 & 30.03 & 59.54 
& 36.08 & - 
& 53.68  & 21.51 & 53.21 & 33.78 & 40.55 & - \\

\noalign{\vskip 1pt}
\hdashline[1pt/3pt]
\noalign{\vskip 1pt}

&Base+I & 56.09 & 55.99 & 55.99 & 62.55 
& 57.66 & \color{deepgreen}{($\uparrow$59.80\%)} 
& 50.39 & 26.72 & 57.31 & 41.81 &  44.06 & \color{deepgreen}{($\uparrow$8.66\%)} \\

&Base+P & 62.13 & 60.57 & 60.66 & 65.12  
& 62.12 & \color{deepgreen}{($\uparrow$72.17\%)} 
& \cellcolor{gray!50}\textbf{65.57} & 37.50 & 64.19 & 48.60 & 53.97 & \color{deepgreen}{($\uparrow$33.10\%)} \\

\multirow{-3}{*}{Text}&Base+I+P & \cellcolor{gray!50}\textbf{63.94} & \cellcolor{gray!30}63.31 & \cellcolor{gray!50}\textbf{64.61} & \cellcolor{gray!30}65.63 
& \cellcolor{gray!30}64.37 & \color{deepgreen}{($\uparrow$78.42\%)}
& 60.49  & 36.98 & \cellcolor{gray!30}64.21 & \cellcolor{gray!50}\textbf{53.77} & 53.86 & \color{deepgreen}{($\uparrow$32.85\%)} \\

\noalign{\vskip 1pt}
\hdashline[1pt/3pt]
\noalign{\vskip 1pt}

&Base+B & 36.05 & 37.73 & 37.18 & 59.26  
& 42.56 & \color{deepgreen}{($\uparrow$17.95\%)}
& 57.16 & 19.13 & 54.03 & 38.86 & 42.30 & \color{deepgreen}{($\uparrow$4.32\%)} \\
&Base+A & 52.43 & 50.94 & 51.31 & 63.75 
& 54.61 & \color{deepgreen}{($\uparrow$43.47\%)}
& 63.04 & \cellcolor{gray!50}\textbf{46.20} & 59.26 & 51.84 & \cellcolor{gray!50}\textbf{55.09} & \color{deepgreen}{($\uparrow$35.86\%)} \\

\multirow{-3}{*}{Code}&Base+B+A & 48.25 & 48.28 & 46.73 
& 63.79 & 51.76
 & \color{deepgreen}{($\uparrow$63.79\%)} & 61.35 & 31.15 & 56.47 & 50.38 & 49.84 & \color{deepgreen}{($\uparrow$22.92\%)} \\

\noalign{\vskip 1pt}
\hdashline[1pt/3pt]
\noalign{\vskip 1pt}

Text + Code &Base+I+P+B+A & \cellcolor{gray!30}63.91 & \cellcolor{gray!50}\textbf{63.70} & \cellcolor{gray!30}64.51 & \cellcolor{gray!50}\textbf{66.83} 
& \cellcolor{gray!50}\textbf{64.74} & \color{deepgreen}{($\uparrow$79.93\%)}
& \cellcolor{gray!30}63.16 &  \cellcolor{gray!30}38.67 & \cellcolor{gray!50}\textbf{65.00} & \cellcolor{gray!30}52.56 & \cellcolor{gray!30}54.85 & \color{deepgreen}{($\uparrow$35.28\%)} \\
\midrule
\multicolumn{14}{c}{\textit{Accuracy}} \\
\midrule
& Base & 22.22 & 21.96 & 24.07 & 44.63 
& 28.22 & -
& 41.67 & 45.28 & 40.50 & 45.45 & 43.23 & - \\

\noalign{\vskip 1pt}
\hdashline[1pt/3pt]
\noalign{\vskip 1pt}

& Base+I & 44.58 & 44.26 & 44.54 & 47.73 
& 45.28 & \color{deepgreen}{($\uparrow$60.44\%)}
& 41.61 & 46.67 & 47.59 & 46.29 & 45.54 & \color{deepgreen}{($\uparrow$5.36\%)} \\

& Base+P & \cellcolor{gray!30}49.95 & 48.52 & 49.07 & \cellcolor{gray!50}\textbf{51.21}
& 49.69 & \color{deepgreen}{($\uparrow$76.07\%)}
 & \cellcolor{gray!50}\textbf{53.33}  & 49.07 & \cellcolor{gray!50}\textbf{53.61} & 47.22 & \cellcolor{gray!50}\textbf{50.81} & \color{deepgreen}{($\uparrow$17.54\%)} \\

\multirow{-3}{*}{Text} & Base+I+P & \cellcolor{gray!50}\textbf{50.19} & \cellcolor{gray!50}\textbf{49.44} & \cellcolor{gray!50}\textbf{51.11} & \cellcolor{gray!30}51.11 
& \cellcolor{gray!50}\textbf{50.46} & \color{deepgreen}{($\uparrow$78.82\%)}
& 47.87  & \cellcolor{gray!50}\textbf{50.42} & \cellcolor{gray!30}52.41 & \cellcolor{gray!50}\textbf{49.54} & \cellcolor{gray!30}50.06 & \color{deepgreen}{($\uparrow$15.81\%)} \\

\noalign{\vskip 1pt}
\hdashline[1pt/3pt]
\noalign{\vskip 1pt}
& Base+B & 35.28 & 36.11 & 35.50 & 44.63 
& 37.88 & \color{deepgreen}{($\uparrow$34.23\%)}
& 42.63  & 46.66 & 42.91 & \cellcolor{gray!30}48.00 &  45.05 & \color{deepgreen}{($\uparrow$4.22\%)} \\

& Base+A & 40.96 & 39.30 & 39.85 & 47.78 
& 41.97 & \color{deepgreen}{($\uparrow$48.73\%)}
& 47.13  & 48.89 & 47.69 & 46.57 &  47.57 & \color{deepgreen}{($\uparrow$10.05\%)} \\

\multirow{-3}{*}{Code} & Base+B+A & 41.02 & 40.22 & 38.98 & 48.70  
& 42.23 & \color{deepgreen}{($\uparrow$49.64\%)}
& 46.11 & 45.45 & 44.81 & 46.10 &  45.62 & \color{deepgreen}{($\uparrow$5.53\%)} \\
\noalign{\vskip 1pt}
\hdashline[1pt/3pt]
\noalign{\vskip 1pt}

 Text + Code & Base+I+P+B+A & 49.91 & \cellcolor{gray!30}49.35 & \cellcolor{gray!30}50.28 & 51.02 
 & \cellcolor{gray!30}50.14 & \color{deepgreen}{($\uparrow$77.68\%)}
 & \cellcolor{gray!30}48.80 & \cellcolor{gray!30}50.37 & 51.94 & 47.54 & 49.66 & \color{deepgreen}{($\uparrow$14.89\%)} \\

\midrule
\multicolumn{14}{c}{\textit{Precision}} \\
\midrule
& Base & 25.88 & 25.41 & 27.85 & 46.91 
& 31.51 & - 
& 44.51 & 38.03 & 43.82 & 42.98 & 42.34 & - \\

\noalign{\vskip 1pt}
\hdashline[1pt/3pt]
\noalign{\vskip 1pt}

& Base+I & 46.47 & 46.26 & 46.41 & 48.76 
& 46.98 & \color{deepgreen}{($\uparrow$49.07\%)}
& 43.84 & 42.68 & 48.35 & 45.51 & 45.10 & \color{deepgreen}{($\uparrow$6.52\%)} \\

& Base+P & 49.94 & 49.08 & 49.42 & \cellcolor{gray!50}\textbf{50.72} 
& 49.79 & \color{deepgreen}{($\uparrow$58.00\%)}
& \cellcolor{gray!50}\textbf{51.95}  & 48.53 & \cellcolor{gray!50}\textbf{52.27} & 47.28 & \cellcolor{gray!30}50.01 & \color{deepgreen}{($\uparrow$18.12\%)} \\

\multirow{-3}{*}{Text} & Base+I+P & \cellcolor{gray!50}\textbf{50.11} & \cellcolor{gray!50}\textbf{49.68} & \cellcolor{gray!50}\textbf{50.63} & \cellcolor{gray!30}50.60 
& \cellcolor{gray!50}\textbf{50.26} & \color{deepgreen}{($\uparrow$59.48\%)}
& 48.70 & \cellcolor{gray!50}\textbf{50.65} & \cellcolor{gray!30}51.45 & \cellcolor{gray!50}\textbf{49.61} & \cellcolor{gray!50}\textbf{50.10} & \color{deepgreen}{($\uparrow$18.35\%)} \\

\noalign{\vskip 1pt}
\hdashline[1pt/3pt]
\noalign{\vskip 1pt}

& Base+B & 35.62 & 36.80 & 36.27 & 46.88 
& 38.89 & \color{deepgreen}{($\uparrow$23.42\%)}
& 45.64 & 39.31 & 45.25 & 47.21 &  44.35 & \color{deepgreen}{($\uparrow$4.77\%)} \\

& Base+A & 43.88 & 42.71 & 43.13 & 48.82 
& 44.64 & \color{deepgreen}{($\uparrow$41.64\%)}
& 48.46  & 48.77 & 48.52 & 47.18 & 48.23 & \color{deepgreen}{($\uparrow$13.93\%)} \\
\multirow{-3}{*}{Code} & Base+B+A & 42.98 & 42.57 & 41.46 & 49.29 
& 44.08 & \color{deepgreen}{($\uparrow$39.87\%)}
& 47.83 & 42.09 & 46.62 & 46.60 &  45.79 & \color{deepgreen}{($\uparrow$8.15\%)} \\

\noalign{\vskip 1pt}
\hdashline[1pt/3pt]
\noalign{\vskip 1pt}

Text + Code & Base+I+P+B+A & \cellcolor{gray!30}49.95 & \cellcolor{gray!30}49.64 & \cellcolor{gray!30}50.15 & 50.52 
& \cellcolor{gray!30}50.07 & \color{deepgreen}{($\uparrow$58.87\%)}
& \cellcolor{gray!30}49.32 & \cellcolor{gray!30}50.60 & 51.11 & \cellcolor{gray!30}48.01 & 49.76 & \color{deepgreen}{($\uparrow$17.54\%)} \\

\bottomrule
\end{tabular}
}
\vspace{-1.6em}
\setlength{\tabcolsep}{6pt} 
\end{table*}

\renewcommand{\arraystretch}{1.0}

\subsection{RQ2: Influence of contexts on effectiveness}
\label{rq2}
In this research question, we investigate the influence of various contexts on model performance in each of the three tasks.

\subsubsection{Influence of context on quality estimation.}

We evaluate the effect of context on the quality estimation task, with the results shown in Table~\ref{tab:task1_result}. We observe that providing context yields substantial performance improvements across all models.
From the table, 
we find that adding textual context, such as the issue description (I) and PR description (P), provides the most boost. For example, considering the closed-source LLMs, adding just the PR description (Base+P) increases the F1-score (Avg$_1$) by 72.17\% (from 36.08 to 62.12). Combining both issue and PR descriptions (Base+I+P) results in the highest F1-score (Avg$_1$) of 64.74, a 79.93\% improvement over the baseline. 
This demonstrates that providing context about the ``true intent behind the code change'' is critical for model performance.
In contrast, the inclusion of only code context (Base+B+A) offers a more modest gain of 63.79\%. This suggests that for a high-level reasoning task like quality estimation, understanding the developer's intent and the problem's background, as articulated in textual descriptions, is more critical than having the full code of the surrounding functions.

\subsubsection{Influence of context on defect localization.}
The results for the defect localization task are presented in Table~\ref{tab:task2_result}. It reveals a more complex and nuanced relationship between context and performance.
Here, a clear divergence emerges between closed-source and open-source models. For closed-source models, adding context generally improves their localization ability. For example, providing the issue description (Base+I) increases their average Pass@1 score (Avg$_1$) by 14.99\%. These models appear capable of effectively parsing the additional information to better localize defects.
Conversely, for open-source models, the inclusion of extra context generally leads to a degradation in performance. Providing the issue description (Base+I) causes a 1.05\% decrease in their average Pass@1 score (Avg$_2$). Similarly, adding code context (Base+B+A) reduces their average Pass@5 score by 3.58\%, while Avg$_1$ increases by 2.53\%. This suggests that the current generation of open-source code models may be more susceptible to the noise introduced by longer contexts, causing their attention to be diffused and their precision to suffer. This highlights a key difference in the contextual reasoning capabilities between the two model classes for this specific task, with closed-source models demonstrating superior performance, likely due to their stronger reasoning capabilities.

\definecolor{deepgreen}{RGB}{0,100,0}
\renewcommand{\arraystretch}{0.8}

\begin{table*}[hbtp]
\setlength{\tabcolsep}{3.5pt} 
\scriptsize 
\centering
\caption{Results in the defect localization task.}
\vspace{-1em}
\label{tab:task2_result}
\footnotesize
\resizebox{\textwidth}{!}{
\begin{tabular}{cccccccccccccc}
\toprule
\rowcolor{gray!20} \textbf{Context} & \textbf{Method} & \textbf{GPT-3.5} & \textbf{GPT-4} & \textbf{GPT-4o} & \textbf{Cl-S} & \color{blue!80}\textbf{Avg$_1$} & \color{blue!80}\textbf{Improve$_1$} & \textbf{Codestral} & \textbf{CQ-32B} & \textbf{CQ-14B} & \textbf{CQ-7B} & \color{blue!80}\textbf{Avg$_2$} & \color{blue!80}\textbf{Improve$_2$}\\
\midrule
\multicolumn{14}{c}{\textit{Pass@1}} \\
\midrule
& Base & 15.74 & 16.48 & 15.83 & 17.59 & 16.41 & - & 17.59 & 24.07 & \cellcolor{gray!50}\textbf{13.70} & \cellcolor{gray!50}\textbf{24.81} & \cellcolor{gray!50}\textbf{20.05} & - \\
\hdashline[1pt/3pt]
& Base+I & \cellcolor{gray!30}18.70 & \cellcolor{gray!50}\textbf{18.43} & \cellcolor{gray!50}\textbf{18.43} & \cellcolor{gray!50}\textbf{19.91} & \cellcolor{gray!50}\textbf{18.87} & \color{deepgreen}{($\uparrow$14.99\%)} & \cellcolor{gray!50}\textbf{19.44} & \cellcolor{gray!30}24.17 & 13.43 & 22.31 & \cellcolor{gray!30}19.84 & \color{red}{($\downarrow$1.05\%)} \\
& Base+P & 16.20 & 16.30 & 15.83 & 18.15 & 16.62 & \color{deepgreen}{($\uparrow$1.28\%)} & 17.78 & \cellcolor{gray!50}\textbf{24.26} & \cellcolor{gray!30}13.61 & \cellcolor{gray!30}23.33 & 19.75 & \color{red}{($\downarrow$1.50\%)} \\
\multirow{-3}{*}{Text} & Base+I+P & \cellcolor{gray!50}\textbf{18.80} & \cellcolor{gray!30}18.06 & \cellcolor{gray!50}\textbf{18.43} & 18.89 & \cellcolor{gray!30}18.54 & \color{deepgreen}{($\uparrow$12.98\%)} & \cellcolor{gray!30}19.26 & 23.89 & 13.43 & 21.85 & 19.61 & \color{red}{($\downarrow$2.19\%)} \\
\hdashline[1pt/3pt]
& Base+B & 15.28 & 14.91 & 15.74 & \cellcolor{gray!30}19.35 & 16.32 & \color{red}{($\downarrow$0.55\%)} & 18.43 & 24.07 & 12.41 & 22.22 & 19.28 & \color{red}{($\downarrow$3.84\%)} \\
& Base+A & 15.83 & 15.93 & 15.56 & 18.80 & 16.53 & \color{deepgreen}{($\uparrow$0.73\%)} & 16.76 & 23.52 & 11.94 & 21.02 & 18.31 & \color{red}{($\downarrow$8.68\%)} \\
\multirow{-3}{*}{Code} & Base+B+A & 15.74 & 15.65 & 14.91 & 17.59 & 15.97 & \color{red}{($\downarrow$2.68\%)} & 18.15 & 23.61 & 12.13 & 20.83 & 18.68 & \color{red}{($\downarrow$6.83\%)} \\
\noalign{\vskip 1pt}
\hdashline[1pt/3pt]
\noalign{\vskip 1pt}
Text + Code & Base+I+P+B+A & 17.13 & 16.94 & 16.57 & 18.80 & 17.36 & \color{deepgreen}{($\uparrow$5.79\%)} & 17.04 & 23.52 & 12.22 & 21.02 & 18.45 & \color{red}{($\downarrow$7.98\%)} \\
\midrule
\multicolumn{14}{c}{\textit{Pass@3}} \\
\midrule
& Base & 34.63 & 34.26 & 33.06 & 39.44 & 35.35 & - & 34.44 & \cellcolor{gray!30}40.83 & \cellcolor{gray!50}\textbf{32.96} & \cellcolor{gray!50}\textbf{36.39} & \cellcolor{gray!30}36.16 & - \\
\hdashline[1pt/3pt]
& Base+I & \cellcolor{gray!50}\textbf{37.13} & 36.39 & \cellcolor{gray!30}37.87 & \cellcolor{gray!50}\textbf{40.19} & \cellcolor{gray!50}\textbf{37.89} & \color{deepgreen}{($\uparrow$7.18\%)} & 35.74 & 40.19 & 31.48 & 35.37 & 35.69 & \color{red}{($\downarrow$1.30\%)} \\
& Base+P & 34.81 & 35.93 & 36.57 & \cellcolor{gray!30}39.54 & 36.71 & \color{deepgreen}{($\uparrow$3.85\%)} & \cellcolor{gray!30}36.02 & 40.28 & \cellcolor{gray!50}\textbf{32.96} & \cellcolor{gray!30}35.46 & \cellcolor{gray!50}\textbf{36.18} & \color{deepgreen}{($\uparrow$0.06\%)} \\
\multirow{-3}{*}{Text} & Base+I+P & \cellcolor{gray!30}36.67 & \cellcolor{gray!30}36.57 & 37.04 & 39.26 & 37.38 & \color{deepgreen}{($\uparrow$5.74\%)} & 35.37 & 39.63 & 31.94 & 34.44 & 35.35 & \color{red}{($\downarrow$2.24\%)} \\
\hdashline[1pt/3pt]
& Base+B & 34.44 & 34.26 & 34.35 & 39.35 & 35.60 & \color{deepgreen}{($\uparrow$0.71\%)} & \cellcolor{gray!50}\textbf{36.39} & 40.28 & 32.41 & 34.35 & 35.86 & \color{deepgreen}{($\uparrow$0.83\%)} \\
& Base+A & 34.54 & 34.91 & 34.63 & 37.78 & 35.46 & \color{red}{($\downarrow$0.31\%)} & 34.54 & 39.44 & 30.74 & 34.07 & 34.70 & \color{red}{($\downarrow$4.04\%)} \\
\multirow{-3}{*}{Code} & Base+B+A & 34.44 & 34.54 & 35.00 & 37.41 & 35.35 & (0.00\%) & 34.63 & \cellcolor{gray!30}40.83 & 31.30 & 32.96 & 34.93 & \color{red}{($\downarrow$3.40\%)} \\
\hdashline[1pt/3pt]
Text + Code & Base+I+P+B+A & 35.83 & \cellcolor{gray!50}\textbf{37.96} & \cellcolor{gray!50}\textbf{38.06} & \cellcolor{gray!30}39.54 & \cellcolor{gray!30}37.85 & \color{deepgreen}{($\uparrow$7.07\%)} & 34.54 & \cellcolor{gray!50}\textbf{41.11} & 31.57 & 33.33 & 35.14 & \color{red}{($\downarrow$2.82\%)} \\
\midrule
\multicolumn{14}{c}{\textit{Pass@5}} \\
\midrule
& Base & 45.09 & 44.81 & 43.43 & 50.00 & 45.83 & - & 44.81 & \cellcolor{gray!30}51.48 & \cellcolor{gray!30}42.59 & \cellcolor{gray!50}\textbf{44.44} & \cellcolor{gray!50}\textbf{45.83} & - \\
\hdashline[1pt/3pt]
& Base+I & \cellcolor{gray!30}47.04 & 47.31 & 46.67 & 49.72 & 47.69 & \color{deepgreen}{($\uparrow$4.06\%)} & 45.37 & 50.19 & 41.76 & 42.87 & 45.05 & \color{red}{($\downarrow$1.70\%)} \\
& Base+P & 46.11 & 46.57 & 47.59 & 49.81 & 47.52 & \color{deepgreen}{($\uparrow$3.69\%)} & \cellcolor{gray!30}45.74 & 50.83 & \cellcolor{gray!50}\textbf{42.78} & \cellcolor{gray!30}42.96 & \cellcolor{gray!30}45.58 & \color{red}{($\downarrow$0.55\%)} \\
\multirow{-3}{*}{Text} & Base+I+P & \cellcolor{gray!50}\textbf{47.13} & \cellcolor{gray!30}47.87 & \cellcolor{gray!50}\textbf{48.98} & 49.91 & \cellcolor{gray!50}\textbf{48.47} & \color{deepgreen}{($\uparrow$5.76\%)} & \cellcolor{gray!30}45.74 & 50.19 & 42.41 & 42.50 & 45.21 & \color{red}{($\downarrow$1.35\%)} \\
\hdashline[1pt/3pt]
& Base+B & 45.46 & 44.81 & 45.19 & \cellcolor{gray!30}50.37 & 46.46 & \color{deepgreen}{($\uparrow$1.37\%)} & \cellcolor{gray!50}\textbf{46.02} & 51.57 & 42.50 & 41.76 & 45.46 & \color{red}{($\downarrow$0.81\%)} \\
& Base+A & 45.74 & 46.20 & 45.65 & 49.72 & 46.83 & \color{deepgreen}{($\uparrow$2.18\%)} & 43.24 & 49.81 & 40.28 & 42.59 & 43.98 & \color{red}{($\downarrow$4.04\%)} \\
\multirow{-3}{*}{Code} & Base+B+A & 45.83 & 46.85 & 45.93 & 49.35 & 46.99 & \color{deepgreen}{($\uparrow$2.53\%)} & 43.43 & \cellcolor{gray!50}\textbf{52.13} & 40.46 & 40.74 & 44.19 & \color{red}{($\downarrow$3.58\%)} \\
\hdashline[1pt/3pt]
Text + Code & Base+I+P+B+A & 46.20 & \cellcolor{gray!50}\textbf{48.33} & \cellcolor{gray!30}47.87 & \cellcolor{gray!50}\textbf{51.11} & \cellcolor{gray!30}48.38 & \color{deepgreen}{($\uparrow$5.56\%)} & 44.81 & 50.74 & 42.31 & 40.09 & 44.49 & \color{red}{($\downarrow$2.92\%)} \\
\midrule
\multicolumn{14}{c}{\textit{Pass@10}} \\
\midrule
& Base & 59.81 & 61.67 & 59.54 & 67.04 & 62.01 & - & 59.72 & 66.85 & \cellcolor{gray!30}60.28 & 55.37 & \cellcolor{gray!30}60.56 & - \\
\hdashline[1pt/3pt]
& Base+I & \cellcolor{gray!50}\textbf{62.96} & 61.76 & 60.37 & 65.93 & 62.75 & \color{deepgreen}{($\uparrow$1.19\%)} & 59.17 & 65.74 & 59.26 & \cellcolor{gray!50}\textbf{57.13} & 60.32 & \color{red}{($\downarrow$0.40\%)} \\
& Base+P & 60.65 & 62.13 & \cellcolor{gray!30}63.52 & \cellcolor{gray!50}\textbf{68.80} & 63.77 & \color{deepgreen}{($\uparrow$2.84\%)} & \cellcolor{gray!50}\textbf{61.57} & \cellcolor{gray!30}67.22 & \cellcolor{gray!50}\textbf{60.93} & 55.74 & \cellcolor{gray!50}\textbf{61.37} & \color{deepgreen}{($\uparrow$1.34\%)} \\
\multirow{-3}{*}{Text} & Base+I+P & \cellcolor{gray!30}61.67 & \cellcolor{gray!30}62.22 & \cellcolor{gray!50}\textbf{64.26} & 68.06 & \cellcolor{gray!50}\textbf{64.05} & \color{deepgreen}{($\uparrow$3.29\%)} & \cellcolor{gray!30}60.28 & 64.72 & 60.00 & \cellcolor{gray!30}56.94 & 60.49 & \color{red}{($\downarrow$0.16\%)} \\
\hdashline[1pt/3pt]
& Base+B & 60.56 & 60.19 & 59.35 & 66.76 & 61.71 & \color{red}{($\downarrow$0.48\%)} & 59.35 & 65.19 & 60.19 & 55.46 & 60.05 & \color{red}{($\downarrow$0.84\%)} \\
& Base+A & 60.00 & 62.04 & 60.56 & 67.31 & 62.48 & \color{deepgreen}{($\uparrow$0.76\%)} & 57.87 & 64.35 & 57.69 & 56.30 & 59.05 & \color{red}{($\downarrow$2.49\%)} \\
\multirow{-3}{*}{Code} & Base+B+A & 60.93 & 60.19 & 61.11 & 67.04 & 62.31 & \color{deepgreen}{($\uparrow$0.48\%)} & 57.69 & 65.93 & 58.70 & 53.80 & 59.03 & \color{red}{($\downarrow$2.53\%)} \\
\hdashline[1pt/3pt]
Text + Code & Base+I+P+B+A & 60.46 & \cellcolor{gray!50}\textbf{62.69} & 63.06 & \cellcolor{gray!30}68.43 & \cellcolor{gray!30}63.66 & \color{deepgreen}{($\uparrow$2.66\%)} & 59.17 & \cellcolor{gray!50}\textbf{67.59} & 59.44 & 54.26 & 60.12 & \color{red}{($\downarrow$0.73\%)} \\
\bottomrule
\end{tabular}
}
\vspace{-1.5em}
\setlength{\tabcolsep}{6pt} 
\end{table*}

\subsubsection{Influence of context on review comment generation.}
For the review comment generation task, we observe consistent but modest improvements from added context, as detailed in Table~\ref{tab:task3_result}.
Across all models, enriching the input with context helps in generating comments that are more aligned with the ground truth. Combining issue and PR descriptions (Base+I+P) yields one of the best results, improving the average Rouge-1 score by 5.73\% for Avg$_1$ and 8.50\% for Avg$_2$, and improving the Edit Similarity by 1.90\% for Avg$_1$ and 4.45\% for Avg$_2$. Unlike in defect localization, both code context and textual context contribute positively, although textual context generally provides a slightly larger benefit. For example, considering closed-source LLMs, adding issue and PR descriptions (Base+I+P) improves Rouge-1 by 5.73\%, while adding before-and-after code (Base+I+P+B+A) improves it by 6.15\%. This indicates that while understanding intent (from text) is important, having the full function scope (from code) also helps the model generate more complete and accurate comments.

\renewcommand{\arraystretch}{0.8}

\begin{table*}[hbtp]
\setlength{\tabcolsep}{3.5pt} 
\scriptsize 
\centering
\caption{Results in the review comment generation task.}
\vspace{-1em}
\label{tab:task3_result}
\footnotesize
\resizebox{\textwidth}{!}{
\begin{tabular}{cccccccccccccccc}
\toprule
\rowcolor{gray!20} \textbf{Context} & \textbf{Method} & \textbf{GPT-3.5} & \textbf{GPT-4} & \textbf{GPT-4o} & \textbf{Cl-S} & \color{blue!80}\textbf{Avg$_1$} & \color{blue!80}\textbf{Improve$_1$} & \textbf{Codestral} & \textbf{CQ-32B} & \textbf{CQ-14B} & \textbf{CQ-7B} & \color{blue!80}\textbf{Avg$_2$} & \color{blue!80}\textbf{Improve$_2$} \\
\midrule
\multicolumn{14}{c}{\textit{ROUGE-1}} \\
\midrule
& Base & 17.93 & 18.11 & 18.16 & 16.69 & 17.72 & - & 12.95 & 16.20 & 16.22 & 13.43 & 14.70 & - \\
\noalign{\vskip 1pt}
\hdashline[1pt/3pt]
\noalign{\vskip 1pt}
& Base+I & \cellcolor{gray!50}\textbf{19.31} & 19.01 & 19.01 & \cellcolor{gray!50}\textbf{17.74} & 18.77 & \color{deepgreen}{($\uparrow$5.90\%)} & \cellcolor{gray!30}14.15 & \cellcolor{gray!30}17.43 & 17.53 & \cellcolor{gray!50}\textbf{14.67} & \cellcolor{gray!50}\textbf{15.95} & \color{deepgreen}{($\uparrow$8.47\%)} \\
& Base+P & 18.70 & 18.62 & 19.02 & 17.36 & 18.43 & \color{deepgreen}{($\uparrow$3.96\%)} & 13.04 & 16.96 & 16.45 & 13.68 & 15.03 & \color{deepgreen}{($\uparrow$2.26\%)} \\
\multirow{-3}{*}{Text} & Base+I+P & \cellcolor{gray!30}19.18 & \cellcolor{gray!50}\textbf{19.30} & \cellcolor{gray!50}\textbf{19.26} & 17.21 & \cellcolor{gray!30}18.74 & \color{deepgreen}{($\uparrow$5.73\%)} & \cellcolor{gray!50}\textbf{14.24} & \cellcolor{gray!50}\textbf{17.45} & \cellcolor{gray!30}17.66 & 14.45 & \cellcolor{gray!50}\textbf{15.95} & \color{deepgreen}{($\uparrow$8.50\%)} \\
\noalign{\vskip 1pt}
\hdashline[1pt/3pt]
\noalign{\vskip 1pt}
& Base+B & 18.78 & 18.97 & 18.68 & 16.93 & 18.34 & \color{deepgreen}{($\uparrow$3.48\%)} & 13.05 & 16.52 & 16.65 & 13.73 & 14.99 & \color{deepgreen}{($\uparrow$1.96\%)} \\
& Base+A & 18.31 & 18.37 & 18.17 & 16.80 & 17.91 & \color{deepgreen}{($\uparrow$1.07\%)} & 13.11 & 16.29 & 16.38 & 13.49 & 14.82 & \color{deepgreen}{($\uparrow$0.80\%)} \\
\multirow{-3}{*}{Code} & Base+B+A & 18.42 & 18.17 & 18.34 & 17.42 & 18.09 & \color{deepgreen}{($\uparrow$2.06\%)} & 12.97 & 16.47 & 17.06 & 13.92 & 15.11 & \color{deepgreen}{($\uparrow$2.76\%)} \\
\noalign{\vskip 1pt}
\hdashline[1pt/3pt]
\noalign{\vskip 1pt}
Text + Code& Base+I+P+B+A & 19.15 & \cellcolor{gray!30}19.22 & \cellcolor{gray!30}19.25 & \cellcolor{gray!30}17.63 & \cellcolor{gray!50}\textbf{18.81} & \color{deepgreen}{($\uparrow$6.15\%)} & 13.82 & 17.16 & \cellcolor{gray!50}\textbf{17.92} & \cellcolor{gray!30}14.53 & 15.86 & \color{deepgreen}{($\uparrow$7.87\%)} \\
\midrule
\multicolumn{14}{c}{\textit{ROUGE-L}} \\
\midrule
& Base & 12.03 & 12.10 & 12.12 & 11.56 & 11.95 & - & 9.91 & 11.62 & 11.37 & 9.88 & 10.70 & - \\
\noalign{\vskip 1pt}
\hdashline[1pt/3pt]
\noalign{\vskip 1pt}
& Base+I & \cellcolor{gray!50}\textbf{12.59} & 12.31 & 12.36 & \cellcolor{gray!30}12.05 & 12.33 & \color{deepgreen}{($\uparrow$3.14\%)} & \cellcolor{gray!30}10.59 & \cellcolor{gray!30}12.09 & 12.06 & \cellcolor{gray!50}\textbf{10.47} & \cellcolor{gray!30}11.30 & \color{deepgreen}{($\uparrow$5.68\%)} \\
& Base+P & 12.25 & 12.32 & 12.41 & 11.94 & 12.23 & \color{deepgreen}{($\uparrow$2.32\%)} & 9.92 & 11.97 & 11.53 & 9.92 & 10.84 & \color{deepgreen}{($\uparrow$1.31\%)} \\
\multirow{-3}{*}{Text} & Base+I+P & \cellcolor{gray!30}12.55 & \cellcolor{gray!50}\textbf{12.63} & \cellcolor{gray!50}\textbf{12.56} & 11.88 & \cellcolor{gray!50}\textbf{12.41} & \color{deepgreen}{($\uparrow$3.79\%)} & \cellcolor{gray!50}\textbf{10.77} & \cellcolor{gray!50}\textbf{12.26} & \cellcolor{gray!30}12.10 & \cellcolor{gray!30}10.42 & \cellcolor{gray!50}\textbf{11.39} & \color{deepgreen}{($\uparrow$6.47\%)} \\
\noalign{\vskip 1pt}
\hdashline[1pt/3pt]
\noalign{\vskip 1pt}
& Base+B & 12.35 & 12.49 & 12.34 & 11.78 & 12.24 & \color{deepgreen}{($\uparrow$2.41\%)} & 10.04 & 11.84 & 11.67 & 9.93 & 10.87 & \color{deepgreen}{($\uparrow$1.64\%)} \\
& Base+A & 12.14 & 12.07 & 12.02 & 11.82 & 12.01 & \color{deepgreen}{($\uparrow$0.50\%)} & 10.07 & 11.55 & 11.52 & 9.82 & 10.74 & \color{deepgreen}{($\uparrow$0.42\%)} \\
\multirow{-3}{*}{Code} & Base+B+A & 12.13 & 12.00 & 12.08 & 11.99 & 12.05 & \color{deepgreen}{($\uparrow$0.82\%)} & 9.84 & 11.69 & 11.95 & 10.06 & 10.89 & \color{deepgreen}{($\uparrow$1.78\%)} \\
\noalign{\vskip 1pt}
\hdashline[1pt/3pt]
\noalign{\vskip 1pt}
Text + Code & Base+I+P+B+A & 12.42 & \cellcolor{gray!30}12.52 & \cellcolor{gray!30}12.45 & \cellcolor{gray!50}\textbf{12.24} & \cellcolor{gray!50}\textbf{12.41} & \color{deepgreen}{($\uparrow$3.81\%)} & 10.37 & 11.98 & \cellcolor{gray!50}\textbf{12.24} & 10.35 & 11.24 & \color{deepgreen}{($\uparrow$5.05\%)} \\
\midrule
\multicolumn{14}{c}{\textit{Edit Similarity}} \\
\midrule
& Base & 21.34 & 21.40 & 21.28 & 20.38 & 21.10 & - & 16.99 & 19.76 & 19.76 & 18.61 & 18.78 & - \\
\noalign{\vskip 1pt}
\hdashline[1pt/3pt]
\noalign{\vskip 1pt}
& Base+I & \cellcolor{gray!50}\textbf{21.80} & 21.73 & \cellcolor{gray!30}21.68 & \cellcolor{gray!50}\textbf{20.90} & \cellcolor{gray!50}\textbf{21.53} & \color{deepgreen}{($\uparrow$2.03\%)} & \cellcolor{gray!50}\textbf{17.88} & \cellcolor{gray!50}\textbf{20.48} & \cellcolor{gray!30}20.82 & \cellcolor{gray!50}\textbf{19.60} & \cellcolor{gray!50}\textbf{19.70} & \color{deepgreen}{($\uparrow$4.87\%)} \\
& Base+P & 21.59 & 21.63 & 21.65 & 20.66 & 21.38 & \color{deepgreen}{($\uparrow$1.34\%)} & 17.29 & 20.02 & 20.33 & 19.07 & 19.18 & \color{deepgreen}{($\uparrow$2.12\%)} \\
\multirow{-3}{*}{Text} & Base+I+P & 21.68 & \cellcolor{gray!50}\textbf{21.81} & \cellcolor{gray!50}\textbf{21.84} & 20.67 & 21.50 & \color{deepgreen}{($\uparrow$1.90\%)} & \cellcolor{gray!30}17.75 & \cellcolor{gray!30}20.44 & 20.74 & \cellcolor{gray!30}19.53 & \cellcolor{gray!30}19.62 & \color{deepgreen}{($\uparrow$4.45\%)} \\
\noalign{\vskip 1pt}
\hdashline[1pt/3pt]
\noalign{\vskip 1pt}
& Base+B & 21.55 & 21.69 & 21.56 & 20.43 & 21.31 & \color{deepgreen}{($\uparrow$0.98\%)} & 17.25 & 19.98 & 20.10 & 18.94 & 19.07 & \color{deepgreen}{($\uparrow$1.53\%)} \\
& Base+A & 21.45 & 21.48 & 21.42 & 20.43 & 21.20 & \color{deepgreen}{($\uparrow$0.45\%)} & 17.21 & 19.68 & 20.05 & 18.73 & 18.92 & \color{deepgreen}{($\uparrow$0.73\%)} \\
\multirow{-3}{*}{Code} & Base+B+A & 21.51 & 21.34 & 21.49 & 20.57 & 21.23 & \color{deepgreen}{($\uparrow$0.60\%)} & 17.13 & 19.82 & 20.38 & 18.90 & 19.06 & \color{deepgreen}{($\uparrow$1.48\%)} \\
\noalign{\vskip 1pt}
\hdashline[1pt/3pt]
\noalign{\vskip 1pt}
Text + Code& Base+I+P+B+A & \cellcolor{gray!30}21.76 & \cellcolor{gray!30}21.77 & 21.66 & \cellcolor{gray!50}\textbf{20.90} & \cellcolor{gray!30}21.52 & \color{deepgreen}{($\uparrow$2.00\%)} & 17.50 & 20.34 & \cellcolor{gray!50}\textbf{20.90} & 19.43 & 19.54 & \color{deepgreen}{($\uparrow$4.06\%)} \\
\bottomrule
\end{tabular}
}
\vspace{-1em}
\setlength{\tabcolsep}{6pt} 
\end{table*}

\renewcommand{\arraystretch}{1.0}

\vspace{-1.5em}

\begin{tcolorbox}[myfindingbox]
    \textbf{Finding 2:} Enriching prompts with additional context generally enhances performance across all code review tasks, but its impact is dependent on the task and model type. Textual context provides the most boost, especially for quality estimation.
\end{tcolorbox}

\subsection{RQ3: Performance on different languages}
\begin{figure*}[ht]
	\centering
	\includegraphics[width=.92\textwidth]{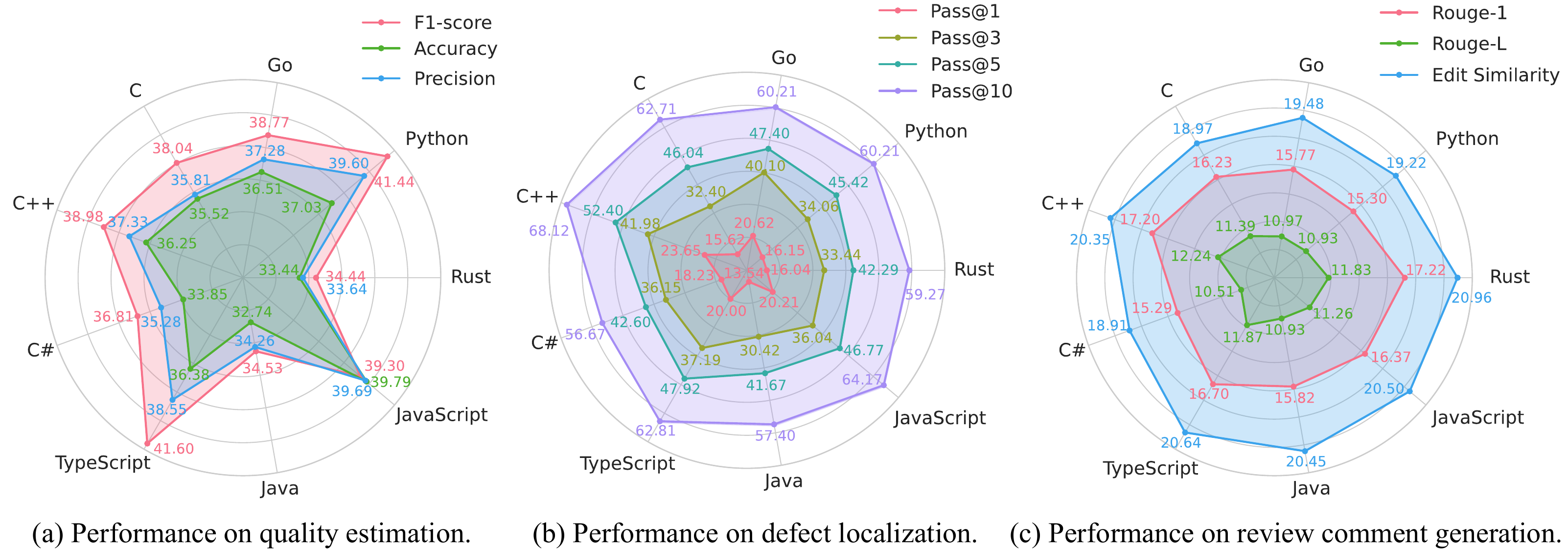}
    \vspace{-1em}
    \caption{Performance across three code review tasks for different programming languages.}
    \vspace{-1em}
\label{fig:performance_language}
\end{figure*}

To understand the influence of programming languages on LLM performance, we experiment with all nine languages across three code review tasks. The performance is detailed in Figure~\ref{fig:performance_language}.
\textbf{(1) LLMs demonstrate overwhelming superiority when analyzing C++ for defect localization.}
As shown in Figure~\ref{fig:performance_language} (b), 
LLMs achieve their highest scores when reviewing C++ code, with the Pass@1 score reaching 23.65 and the Pass@10 score reaching 68.12.
However, this dominance in C++ does not extend to other code review tasks. For instance, LLMs perform best on Python and JavaScript for quality estimation, leading in both F1-score and Accuracy as shown in Figure~\ref{fig:performance_language} (a).
For comment generation, they deliver excellent performance on languages like TypeScript and Rust, as shown in Figure~\ref{fig:performance_language} (c).
These results underscore that, although LLMs demonstrate superior overall performance in certain programming languages, the most effective language for a code review task can differ substantially. Such variability may be attributed to the domain knowledge of programming languages present in the LLMs’ training data.

\textbf{(2) LLMs show 
better performance on web-centric languages like TypeScript and JavaScript, but struggle with Java.}
LLMs demonstrate robust capabilities when reviewing TypeScript and JavaScript, ranking among the top-performing languages across all three tasks.
For example, as illustrated in Figure~\ref{fig:performance_language} (c), in the comment generation task for TypeScript, LLMs achieve high scores on both Edit Similarity (20.64) and Rouge-L (11.87).
In addition, LLMs consistently exhibit lower performance when reviewing Java code. This trend is evident in that Java-related data points are predominantly situated in the inner regions. For example, LLMs achieve one of the lowest F1-scores (34.53) in quality estimation and a comparatively low Edit Similarity (20.45) in comment generation tasks. This may be attributable to the more intricate class and method dependencies that are characteristic of Java projects.


\begin{tcolorbox}[myfindingbox]
    \textbf{Finding 3:} LLM performance is greatly influenced by programming language. C++ shows exceptional performance in defect localization, outperforming all other languages on every metric for that task. Web languages like TypeScript and JavaScript demonstrate strong, balanced capabilities, while Java's performance is generally among the weakest across all tasks.
\end{tcolorbox}

\section{Discussion}
\label{discussion}
\begin{figure}[ht]
	\centering
	\includegraphics[width=.47\textwidth]{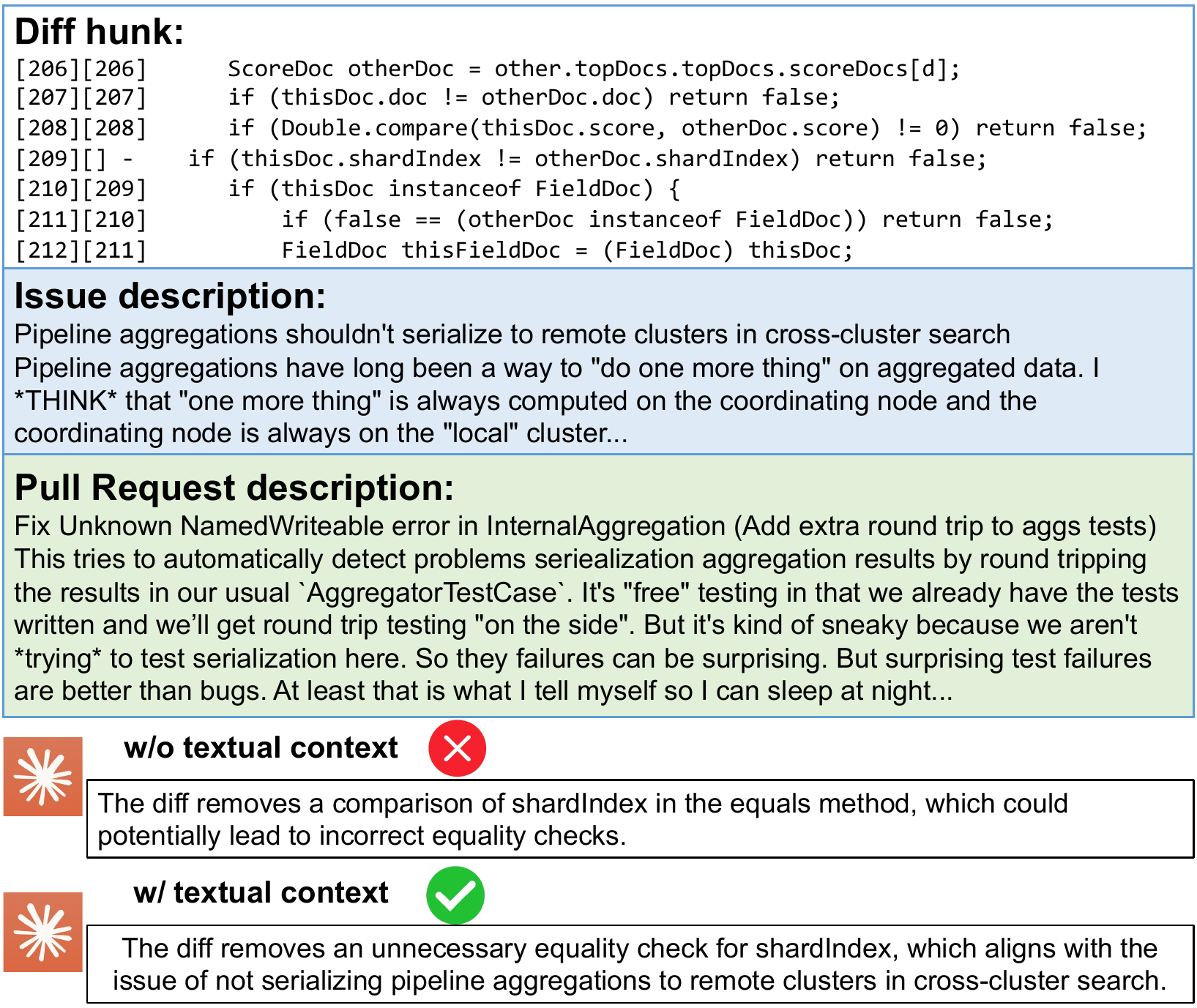}
    \vspace{-1em}
    \caption{A case study demonstrating the positive impact of textual context.}
    \vspace{-2.2em}
\label{fig:case_study}
\end{figure}

\subsection{Industry Application at ByteDance}
\begin{table}[t]
\setlength{\tabcolsep}{8pt} 
  \centering
  \caption{Performance improvements of ByteDance's internal code review tool by leveraging \dataset. The tasks are Quality Estimation (QE), Defect Localization (DL), and Review Comment Generation (RCG).}
\vspace{-1em}
  
  \label{tab:industry}
  \begin{tabular}{lccccc}
    \toprule
    \rowcolor{gray!20} & \textbf{QE} & \textbf{DL} & \textbf{RCG} & \color{blue!80}\textbf{Avg} & \color{blue!80}\textbf{Improve}\\
    \midrule
    Before & 20.00 & 28.89 & 22.35 & 23.75 & - \\
    After  & \textbf{56.00} & \textbf{33.33} & \textbf{26.07} & \textbf{38.47} & \color{deepgreen}{($\uparrow$61.98\%)}\\
    \bottomrule
  \end{tabular}
\vspace{-1.5em}
\end{table}
\dataset's practical value has been demonstrated in a real-world application by a team at ByteDance. We aim to build a high-quality code review tool using a self-iteration methodology. The process begins with an initial version of the tool generated by Claude 4. Then, this tool is refined through an iterative loop designed to optimize its context extraction and prompt engineering capabilities.
In each round of this self-iteration, \dataset is used as the authoritative evaluation mechanism. It assesses the tool's quality by calculating performance scores across our three core tasks: quality estimation, defect localization, and comment generation. These scores, derived from the metrics detailed in Section~\ref{metric}, serve as the crucial reward signal that guides the optimization process.

After 100 rounds of iteration, we assess the tool's initial and final versions. To avoid data leakage, we use a reserved set of 25 repositories from our benchmark. The results are shown in the Table~\ref{tab:industry}, which shows a great improvement: the average final score surges from 0.2375 to 0.3847, marking a 61.98\% relative improvement. This demonstrates that \dataset is not just an academic benchmark; it is a practical and effective tool for rigorously evaluating and quantifiably driving the advancement of sophisticated code review systems in an industrial setting.

\subsection{Case Study}
To further understand how the textual context provided by \dataset enhances the performance of LLMs in code review tasks, we present an example in Figure~\ref{fig:case_study}.
Specifically, we employ Claude 3.5 Sonnet to evaluate a diff hunk from the repository ``\texttt{elastic/elasticsearch}'', corresponding to a valid change that was successfully merged. 
When the textual context is omitted, the LLM incorrectly predicts that the change should not be merged.
The model's reasoning is based on a sound but context-free software engineering principle~\cite{bloch2008effective}: by removing the ``\texttt{shardIndex}'' comparison from an ``\texttt{equals()}'' method, the change appears to weaken the method's correctness contract, potentially leading to incorrect equality checks and bugs. 
However, when the model is provided with the relevant issue and PR descriptions, its assessment is completely reversed. 
When provided with the additional context that the modification is an intentional bug fix addressing serialization issues in cross-cluster search, where ``\texttt{shardIndex}'' is inherently transient, the LLM correctly determines that the change is appropriate for merging. The model recognizes that the removal of the check is not only justified, but is in fact the primary objective of the fix.
This indicates that textual context plays a critical role in guiding the model’s code review process.

\subsection{Implication of Findings}
Based on our findings, we provide the following implications for developers and researchers who use LLMs for code review.

\textbf{Implications for Developers.}
As LLM-based code review tools are increasingly adopted in software development workflows, developers face the challenge of selecting appropriate context to use as prompts.  Our findings indicate that the most effective way to improve the quality of LLM-generated code reviews is to provide high-quality textual context, such as detailed PR descriptions. By prioritizing clear and comprehensive descriptions that articulate the intent and rationale behind code changes, LLMs can effectively enhance code review outcomes. It also underscores the necessity for organizations to enforce more rigorous standards for PR and issue documentation. Such practices not only optimize the performance of AI tools but also greatly improve overall project management and knowledge sharing within teams.

\textbf{Implications for Researchers.}
We identify a gap between understanding code syntax and comprehending its semantic intent. This indicates that current LLMs often struggle to capture the underlying rationale or the reason behind code changes, which is essential for high-quality code review. Our findings indicate that this challenge cannot be effectively addressed simply by increasing model size or by indiscriminately adding more context. In some cases, open-source models may even be negatively affected by excessive or irrelevant context. We recommend that LLM researchers focus on improving how models utilize textual context, for example, by developing training methods that teach models to prioritize information from issue descriptions and PR summaries over the code itself. Furthermore, future research should explore techniques that enhance models' ability to infer and reason about the intent behind code changes, thereby bridging the gap between syntactic understanding and semantic comprehension.

\subsection{Threats to Validity}

We identify two threats to the validity of our study. First, our benchmark's scope is limited to nine mainstream programming languages, which may restrict the generalizability of our findings. However, our benchmark construction pipeline is extensible. In future research, we intend to extend our investigations to more programming languages. Second, our evaluation is confined to a representative set of LLMs~\cite{hou2024large} due to computational costs. This set includes leading proprietary models, such as the GPT series, and state-of-the-art open-source models, like the Qwen series, ensuring our findings are relevant to the current state of the field.


\section{Related Work}
\label{relatedwork}
\subsection{Code Review Benchmark}

While researchers have constructed a variety of code review benchmarks in recent years, these benchmarks still face challenges. For instance, Li et al.\cite{10.1145/3540250.3549081} and Hong et al.\cite{9825760} provide only code diff hunks, lacking issue or pull request descriptions needed for understanding change motivation. Even when PR descriptions are included by Rong et al. and CORRECT~\cite{7883306, 9793876}, the deeper rationale in linked issues is often missing. Data quality is also a concern. Benchmarks like Li et al.\cite{10.1145/3540250.3549081}, CORMS\cite{10.1145/3540250.3549115}, and Tufano et al.\cite{10.1145/3510003.3510621} mainly use basic filtering, rarely removing outdated or irrelevant comments. Besides, most benchmarks\cite{10.1145/3540250.3549081, 10.1145/3540250.3549115, paixao2018crop} support only coarse-grained, hunk-level tasks; even line-level benchmarks~\cite{9825760} are limited to binary classification, missing the richness of review feedback. To address these gaps, we introduce \dataset, which offers comprehensive context, rigorous filtering, and fine-grained line-level annotations for more reliable code review evaluation. 

\subsection{LLMs in Code Review}

With the widespread adoption of LLMs in software engineering~\cite{llm1, llm2}, research on code review automation has made great progress. 
For review comment generation, Moshkin et al.\cite{9896360} develop an automated review tool for Python using a pre-trained neural network, while AUGER\cite{li2022auger} and subsequent works~\cite{10.1145/3540250.3549081, 10.1145/3510003.3510621, zhou2023generation, lin2023towards} leverage and fine-tune models like T5 and CodeT5 to improve comment quality and efficiency across multiple languages and benchmarks. In code refinement, Tufano et al.\cite{10.1145/3510003.3510621}, Li et al.\cite{10.1145/3540250.3549081}, Zhou et al.\cite{zhou2023generation}, and Lin et al.\cite{lin2023towards} demonstrate the effectiveness of LLMs such as CodeBERT and GraphCodeBERT, while Yin et al.\cite{yin2023automatic} further enhance performance by incorporating program dependency graphs. For quality estimation, Li et al.\cite{10.1145/3540250.3549081} and Sawant et al.~\cite{sawant2023code} employ multi-task pre-training and fine-tuning strategies to improve model generalization and support code compliance.


\section{Conclusion}
\label{conclusion}
In this paper, we introduce \dataset, a large-scale benchmark for fine-grained, context-enriched code review, featuring a three-stage evaluation framework. Our experiments reveal two key findings: (1) even state-of-the-art LLMs still face significant limitations in code review, and (2) their performance is substantially enhanced by rich textual context, such as issue descriptions. The practical value of our benchmark is demonstrated through its application at ByteDance, where it guides a self-evolving code review tool to a 61.98\% relative performance improvement by serving as its core reward signal. In conclusion, \dataset provides a robust foundation for developing more accurate and practical automated code review systems.

\bibliographystyle{ACM-Reference-Format}
\bibliography{references}

@inproceedings{paixao2018crop,
  title={CROP: Linking code reviews to source code changes},
  author={Paixao, Matheus and Krinke, Jens and Han, Donggyun and Harman, Mark},
  booktitle={Proceedings of the 15th international conference on mining software repositories},
  pages={46--49},
  year={2018}
}

@INPROCEEDINGS{7883306,
  author={Rahman, Mohammad Masudur and Roy, Chanchal K. and Collins, Jason A.},
  booktitle={2016 IEEE/ACM 38th International Conference on Software Engineering Companion (ICSE-C)}, 
  title={CORRECT: Code Reviewer Recommendation in GitHub Based on Cross-Project and Technology Experience}, 
  year={2016},
  volume={},
  number={},
  pages={222-231},
  doi={}}

@inproceedings{10.1145/3510003.3510621, author = {Tufano, Rosalia and Masiero, Simone and Mastropaolo, Antonio and Pascarella, Luca and Poshyvanyk, Denys and Bavota, Gabriele}, title = {Using pre-trained models to boost code review automation}, year = {2022}, isbn = {9781450392211}, publisher = {Association for Computing Machinery}, address = {New York, NY, USA}, url = {https://doi.org/10.1145/3510003.3510621}, doi = {10.1145/3510003.3510621}, booktitle = {Proceedings of the 44th International Conference on Software Engineering}, pages = {2291–2302}, numpages = {12}, location = {Pittsburgh, Pennsylvania}, series = {ICSE '22} }

@inproceedings{10.1145/3540250.3549081, author = {Li, Zhiyu and Lu, Shuai and Guo, Daya and Duan, Nan and Jannu, Shailesh and Jenks, Grant and Majumder, Deep and Green, Jared and Svyatkovskiy, Alexey and Fu, Shengyu and Sundaresan, Neel}, title = {Automating code review activities by large-scale pre-training}, year = {2022}, isbn = {9781450394130}, publisher = {Association for Computing Machinery}, address = {New York, NY, USA}, url = {https://doi.org/10.1145/3540250.3549081}, doi = {10.1145/3540250.3549081}, booktitle = {Proceedings of the 30th ACM Joint European Software Engineering Conference and Symposium on the Foundations of Software Engineering}, pages = {1035–1047}, numpages = {13}, location = {Singapore, Singapore}, series = {ESEC/FSE 2022} }

@INPROCEEDINGS{9825760,
  author={Hong, Yang and Tantithamthavorn, Chakkrit Kla and Thongtanunam, Patanamon Pick},
  booktitle={2022 IEEE International Conference on Software Analysis, Evolution and Reengineering (SANER)}, 
  title={Where Should I Look at? Recommending Lines that Reviewers Should Pay Attention To}, 
  year={2022},
  volume={},
  number={},
  pages={1034-1045},
  doi={10.1109/SANER53432.2022.00121}}

@INPROCEEDINGS{9896360,
  author={Moshkin, Vadim and Kalachev, Vladislav and Zarubin, Anton},
  booktitle={2022 International Russian Automation Conference (RusAutoCon)}, 
  title={Automation of Program Code Analysis Using Machine Learning Methods}, 
  year={2022},
  volume={},
  number={},
  pages={404-408},
  doi={10.1109/RusAutoCon54946.2022.9896360}}

@inproceedings{li2022auger,
  title={AUGER: automatically generating review comments with pre-training models},
  author={Li, Lingwei and Yang, Li and Jiang, Huaxi and Yan, Jun and Luo, Tiejian and Hua, Zihan and Liang, Geng and Zuo, Chun},
  booktitle={Proceedings of the 30th ACM Joint European Software Engineering Conference and Symposium on the Foundations of Software Engineering},
  pages={1009--1021},
  year={2022}
}

@inproceedings{zhou2023generation,
  title={Generation-based code review automation: How far are we?},
  author={Zhou, Xin and Kim, Kisub and Xu, Bowen and Han, DongGyun and He, Junda and Lo, David},
  booktitle={2023 IEEE/ACM 31st International Conference on Program Comprehension (ICPC)},
  pages={215--226},
  year={2023},
  organization={IEEE}
}

@inproceedings{lin2023towards,
  title={Towards automated code reviews: Does learning code structure help?},
  author={Lin, Hong Yi and Thongtanunam, Patanamon},
  booktitle={2023 IEEE International Conference on Software Analysis, Evolution and Reengineering (SANER)},
  pages={703--707},
  year={2023},
  organization={IEEE}
}

@article{yin2023automatic,
  title={Automatic code review by learning the structure information of code graph},
  author={Yin, Ying and Zhao, Yuhai and Sun, Yiming and Chen, Chen},
  journal={Sensors},
  volume={23},
  number={5},
  pages={2551},
  year={2023},
  publisher={MDPI}
}

@inproceedings{sawant2023code,
  title={Code compliance assessment as a learning problem},
  author={Sawant, Neela and Sengamedu, Srinivasan H},
  booktitle={2023 IEEE/ACM 45th International Conference on Software Engineering: Software Engineering in Practice (ICSE-SEIP)},
  pages={445--454},
  year={2023},
  organization={IEEE}
}

@inproceedings{10.1145/3540250.3549115,
author = {Pandya, Prahar and Tiwari, Saurabh},
title = {CORMS: a GitHub and Gerrit based hybrid code reviewer recommendation approach for modern code review},
year = {2022},
isbn = {9781450394130},
publisher = {Association for Computing Machinery},
address = {New York, NY, USA},
url = {https://doi.org/10.1145/3540250.3549115},
doi = {10.1145/3540250.3549115},
pages = {546–557},
numpages = {12},
location = {Singapore, Singapore},
series = {ESEC/FSE 2022}
}

@INPROCEEDINGS{9793876,
  author={Rong, Guoping and Zhang, Yifan and Yang, Lanxin and Zhang, Fuli and Kuang, Hongyu and Zhang, He},
  booktitle={2022 IEEE/ACM 44th International Conference on Software Engineering (ICSE)}, 
  title={Modeling Review History for Reviewer Recommendation: A Hypergraph Approach}, 
  year={2022},
  volume={},
  number={},
  pages={1381-1392},
  doi={10.1145/3510003.3510213}}

@inproceedings{survey1,
  title={Expectations, outcomes, and challenges of modern code review},
  author={Bacchelli, Alberto and Bird, Christian},
  booktitle={2013 35th International Conference on Software Engineering (ICSE)},
  pages={712--721},
  year={2013},
  organization={IEEE}
}

@article{survey2,
  author       = {Zezhou Yang and
                  Cuiyun Gao and
                  Zhaoqiang Guo and
                  Zhenhao Li and
                  Kui Liu and
                  Xin Xia and
                  Yuming Zhou},
  title        = {A Survey on Modern Code Review: Progresses, Challenges and Opportunities},
  journal      = {CoRR},
  volume       = {abs/2405.18216},
  year         = {2024},
  url          = {https://doi.org/10.48550/arXiv.2405.18216},
  doi          = {10.48550/ARXIV.2405.18216},
  eprinttype    = {arXiv},
  eprint       = {2405.18216},
  bibsource    = {dblp computer science bibliography, https://dblp.org}
}

@article{survey3,
  title={Modern code reviews—survey of literature and practice},
  author={Badampudi, Deepika and Unterkalmsteiner, Michael and Britto, Ricardo},
  journal={ACM Transactions on Software Engineering and Methodology},
  volume={32},
  number={4},
  pages={1--61},
  year={2023},
  publisher={ACM New York, NY, USA}
}

@inproceedings{survey4,
  title={Code review quality: How developers see it},
  author={Kononenko, Oleksii and Baysal, Olga and Godfrey, Michael W},
  booktitle={Proceedings of the 38th international conference on software engineering},
  pages={1028--1038},
  year={2016}
}

@inproceedings{survey5,
  title={Impact of peer code review on peer impression formation: A survey},
  author={Bosu, Amiangshu and Carver, Jeffrey C},
  booktitle={2013 ACM/IEEE International Symposium on Empirical Software Engineering and Measurement},
  pages={133--142},
  year={2013},
  organization={IEEE}
}

@article{llm1,
  title={Harnessing the power of llms in practice: A survey on chatgpt and beyond},
  author={Yang, Jingfeng and Jin, Hongye and Tang, Ruixiang and Han, Xiaotian and Feng, Qizhang and Jiang, Haoming and Zhong, Shaochen and Yin, Bing and Hu, Xia},
  journal={ACM Transactions on Knowledge Discovery from Data},
  volume={18},
  number={6},
  pages={1--32},
  year={2024},
  publisher={ACM New York, NY}
}

@article{llm2,
  title={A survey on llm-generated text detection: Necessity, methods, and future directions},
  author={Wu, Junchao and Yang, Shu and Zhan, Runzhe and Yuan, Yulin and Chao, Lidia Sam and Wong, Derek Fai},
  journal={Computational Linguistics},
  volume={51},
  number={1},
  pages={275--338},
  year={2025},
  publisher={MIT Press 255 Main Street, 9th Floor, Cambridge, Massachusetts 02142, USA~…}
}

@article{llm3,
  title={A survey on large language models for recommendation},
  author={Wu, Likang and Zheng, Zhi and Qiu, Zhaopeng and Wang, Hao and Gu, Hongchao and Shen, Tingjia and Qin, Chuan and Zhu, Chen and Zhu, Hengshu and Liu, Qi and others},
  journal={World Wide Web},
  volume={27},
  number={5},
  pages={60},
  year={2024},
  publisher={Springer}
}

@misc{topprogramminglanguage,
  author = {global-metrics},
  title = {{Programming languages}},
  note = {\url{https://innovationgraph.github.com/global-metrics/programming-languages}},
}

@misc{top100,
  author = {EvanLi},
  title = {{Github-Ranking}},
  note = {\url{https://github.com/EvanLi/Github-Ranking/blob/master/Top100}},
}

@misc{git3,
    author = {GitHub},
    year  = {[n.d.]},
    title = {Using keywords in issues and pull requests},
    note = {\url{https://docs.github.com/en/rest?apiVersion=2022-11-28}},
}

@misc{git2,
  author = {GitHub},
    year  = {[n.d.]},
  title = {{GitHub Graph API}},
  note = {\url{https://docs.github.com/en/graphql}},
}

@misc{Tree-sitter,
    author = {tree-sitter},
    year  = {[n.d.]},
    title = {“{T}ree-sitter”},
    note  = {\url{https://tree-sitter.github.io/tree-sitter/}},
}

@inproceedings{codejudge,
  author       = {Weixi Tong and
                  Tianyi Zhang},
  editor       = {Yaser Al{-}Onaizan and
                  Mohit Bansal and
                  Yun{-}Nung Chen},
  title        = {CodeJudge: Evaluating Code Generation with Large Language Models},
  booktitle    = {Proceedings of the 2024 Conference on Empirical Methods in Natural
                  Language Processing, {EMNLP} 2024, Miami, FL, USA, November 12-16,
                  2024},
  pages        = {20032--20051},
  publisher    = {Association for Computational Linguistics},
  year         = {2024},
  url          = {https://aclanthology.org/2024.emnlp-main.1118},
  bibsource    = {dblp computer science bibliography, https://dblp.org}
}

@inproceedings{ice-score,
  author       = {Terry Yue Zhuo},
  editor       = {Yvette Graham and
                  Matthew Purver},
  title        = {ICE-Score: Instructing Large Language Models to Evaluate Code},
  booktitle    = {Findings of the Association for Computational Linguistics: {EACL}
                  2024, St. Julian's, Malta, March 17-22, 2024},
  pages        = {2232--2242},
  publisher    = {Association for Computational Linguistics},
  year         = {2024},
  url          = {https://aclanthology.org/2024.findings-eacl.148},
  bibsource    = {dblp computer science bibliography, https://dblp.org}
}

@misc{bavaresco2025llmsinsteadhumanjudges,
      title={LLMs instead of Human Judges? A Large Scale Empirical Study across 20 NLP Evaluation Tasks}, 
      author={Anna Bavaresco and Raffaella Bernardi and Leonardo Bertolazzi and Desmond Elliott and Raquel Fernández and Albert Gatt and Esam Ghaleb and Mario Giulianelli and Michael Hanna and Alexander Koller and André F. T. Martins and Philipp Mondorf and Vera Neplenbroek and Sandro Pezzelle and Barbara Plank and David Schlangen and Alessandro Suglia and Aditya K Surikuchi and Ece Takmaz and Alberto Testoni},
      year={2025},
      eprint={2406.18403},
      archivePrefix={arXiv},
      primaryClass={cs.CL},
      url={https://arxiv.org/abs/2406.18403}, 
}

@book{bloch2008effective,
  title={Effective java},
  author={Bloch, Joshua},
  year={2008},
  publisher={Addison-Wesley Professional}
}

@inproceedings{sung2018learning,
  title={Learning to compare: Relation network for few-shot learning},
  author={Sung, Flood and Yang, Yongxin and Zhang, Li and Xiang, Tao and Torr, Philip HS and Hospedales, Timothy M},
  booktitle={Proceedings of the IEEE conference on computer vision and pattern recognition},
  pages={1199--1208},
  year={2018}
}

@article{wei2022chain,
  title={Chain-of-thought prompting elicits reasoning in large language models},
  author={Wei, Jason and Wang, Xuezhi and Schuurmans, Dale and Bosma, Maarten and Xia, Fei and Chi, Ed and Le, Quoc V and Zhou, Denny and others},
  journal={Advances in neural information processing systems},
  volume={35},
  pages={24824--24837},
  year={2022}
}

@article{hou2024large,
  title={Large language models for software engineering: A systematic literature review},
  author={Hou, Xinyi and Zhao, Yanjie and Liu, Yue and Yang, Zhou and Wang, Kailong and Li, Li and Luo, Xiapu and Lo, David and Grundy, John and Wang, Haoyu},
  journal={ACM Transactions on Software Engineering and Methodology},
  volume={33},
  number={8},
  pages={1--79},
  year={2024},
  publisher={ACM New York, NY}
}

\end{document}